\documentclass{aa}  
\usepackage{moresize}
\usepackage{graphicx}
\usepackage{txfonts,textcomp}
\DeclareMathOperator*{\argmin}{arg\,min}

\usepackage{algpseudocode}
\usepackage[linesnumbered,ruled,vlined]{algorithm2e}
\usepackage{bbold}
\usepackage{hyperref}
\usepackage{float}
\usepackage{array}
\usepackage[switch]{lineno}
\usepackage{datetime}
\usepackage[usenames,table,dvipsnames]{xcolor}
\usepackage{setspace}
\usepackage[normalem]{ulem}
\usepackage{soul}
\usepackage{subcaption}
\usepackage{dblfloatfix}

\usepackage{makecell}

\begin{document} 

   \title{Best of both worlds: Fusing  hyperspectral data from two generations of spectro-imagers for X-ray astrophysics}

   \author{J. Lascar 
          \inst{1}
          \and
          J. Bobin \inst{1}
          \and
          F. Acero \inst{1,2}
          }

   \institute{Université Paris-Saclay, Université Paris Cité, CEA, CNRS, AIM, 91191, Gif-sur-Yvette, France \\
              \email{julia.lascar@cea.fr}
              \and
            FSLAC IRL 2009, CNRS/IAC, La Laguna, Tenerife, Spain \\              
             }
 

  \abstract
  {With the recent launch of the X-Ray Imaging and Spectroscopy Mission (XRISM) and the advent of microcalorimeter detectors, X-ray astrophysics is entering in a new era of spatially resolved high-resolution spectroscopy.
  But while this new generation of X-ray telescopes offers much finer spectral resolution than the previous one (e.g. XMM-Newton, Chandra), these instruments also have coarser spatial resolutions, leading to problematic cross-pixel contamination. This issue is currently a critical limitation for the study of extended sources, such as galaxy clusters or supernova remnants (SNRs).}
  {To increase the scientific output of XRISM's hyperspectral data, we propose that it be fused with XMM-Newton data, and seek to obtain a cube with the best spatial and spectral resolution of both generations. 
  This is the aim of hyperspectral fusion. In this article, we explore the potential and limitations of hyperspectral fusion for X-ray astrophysics.}
  {We implemented an algorithm that jointly deconvolves the spatial response of XRISM and the spectral response of XMM-Newton. To do so, we construct a forward model adapted for instrumental systematic degradations and Poisson noise, and then tackle hyperspectral fusion as a regularized inverse problem. We test three methods of regularization: spectral low-rank approximation with a spatial Sobolev regularization; spectral low-rank approximation with a 2D wavelet sparsity constraint; and a 2D-1D wavelet sparsity constraint.} 
  {We test our method on toy models constructed from hydrodynamic simulations of SNRs. We find that our method reconstructs the ground truth well even when the toy model is complex. For the regularization term, we find that while the low-rank approximation works well as a spectral denoiser in models with less spectral variability, it introduces a bias in models with more spectral variability, in which case the 2D-1D wavelet sparsity regularization works best. Following the present proof of concept, we aim to apply this method to real X-ray astrophysical data in the near future.}
{}


   \maketitle
%

\section{Introduction}
\label{sec:intro}
With the advent of microcalorimeter detectors, X-ray astrophysics has entered a new  era of spatially resolved high-resolution spectroscopy. Indeed, the recent launch of the X-Ray Imaging and Spectroscopy Mission (XRISM) and its microcalorimeter array "Resolve"\footnote{\url{https://xrism.isas.jaxa.jp}}, and the future Athena telescope and its X-ray Integral Field Unit (X-IFU) \footnote{\url{https://x-ifu.irap.omp.eu/}} camera, has granted access to spatially resolved spectroscopy at spectral resolutions improved by an order of magnitude. This is set to allow astrophysicists to investigate previously indistinguishable spectral features, such as the emission lines of rare elements, line broadening, or velocity measurements.
These characteristics are essential in the study of extended sources such as supernova remnants (SNRs), galaxy clusters, and many of the other exotic sources of the X-ray Universe. 

However, while the XRISM/Resolve instrument has a much finer spectral resolution than the Chandra X-ray Observatory's Advanced CCD Imaging Spectrometer (Chandra/ACIS)\footnote{\url{https://cxc.harvard.edu/}} or the X-ray Multi-Mirror-Newton's European Photon Imaging Camera (XMM-Newton/EPIC)\footnote{\url{https://www.cosmos.esa.int/web/xmm-newton}} cameras, its spatial resolution is much lower than that of its predecessors. As we demonstrate here, this spatial blur is an important obstacle when analyzing XRISM/Resolve data. Given this trade-off, we set out to establish whether or not it is possible to combine the data from these two generations of X-ray observatories and obtain a result with the best spectral and spatial resolution of each instrument. This is the idea at the heart of hyperspectral fusion.

\begin{figure*}[t]
\begin{subfigure}{\textwidth}
    \centering
    \includegraphics[width=0.6\textwidth]{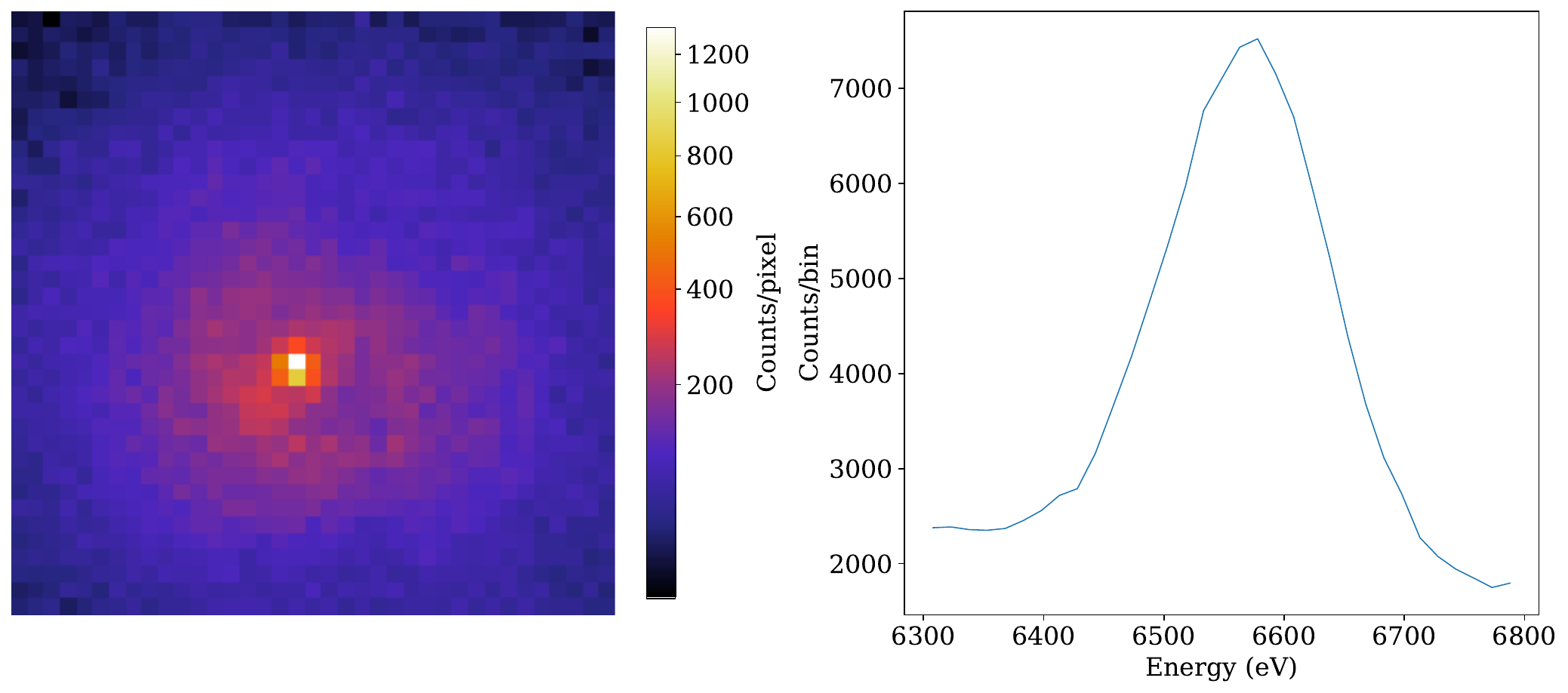}
\end{subfigure}
\begin{subfigure}{\textwidth}
    \centering
    \includegraphics[width=0.6\textwidth]{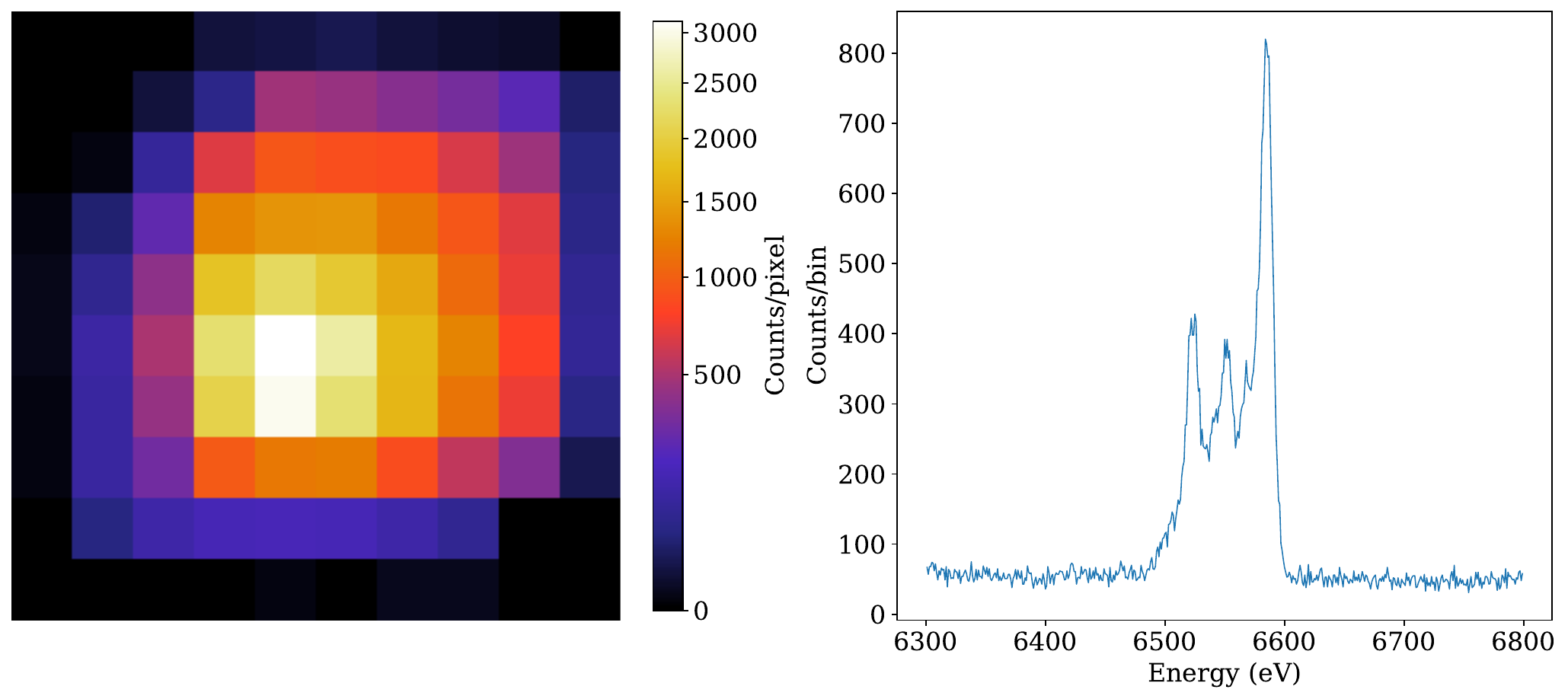}
\end{subfigure}
    \caption{Comparison of hyperspectral data of the Perseus galaxy cluster as observed by two generations of X-ray spectro-imagers.
    The XMM-Newton/MOS (top) and Hitomi/SXS (bottom) data have pixels of 8"/30" and spectral channels of 15/1 eV respectively.
    The left panels show the  counts maps summed over all spectral channels and the right panels the spectra summed over all pixels.}
    \label{fig:Hitomi_XMM}
\end{figure*}

In this paper, we explore the potential and limitations of hyperspectral fusion for X-ray data taken by pairs of telescopes such as XMM-Newton/EPIC with XRISM/Resolve, or Chandra with Athena-X-IFU.  

We  implemented a modular optimization algorithm that can use one of three potential regularizations: first, inspired by \cite{guilloteau_2020}, and given the widespread usage of low-rank approximations in the literature, we implement a low-rank approximation with Sobolev spatial regularization. Since Sobolev regularization does not take into account multi-scale features, we then considered a low-rank approximation with a 2D wavelet sparsity regularization. Finally, to investigate whether low-rankedness is indeed appropriate, we implemented a heuristic approach using a 2D-1D Wavelet sparsity regularization. This hyperspectral fusion code is openly available on Github.\footnote{\url{https://github.com/JMLascar/HIFReD_fusion}}

In Section \ref{sec:context}, we present the astrophysical and methodological context behind our work. In Section \ref{sec:methods} we present our methodology, and describe the forward model we have developed for X-ray astrophysics, and the proposed regularization schemes. In section \ref{sec:results}, we test our method on toy models of varying complexity using synthetic X-ray data cubes constructed from hydrodynamic simulation of SNRs. We discuss the achievements, limitations, and perspective for future work in section \ref{sec:discussion}.
\section{Context}
\label{sec:context}
\subsection{Astrophysical context: X-ray spectro-imaging and microcalorimeters}


X-ray spectro-imagers can collect hyperspectral images. These are 3D data arrays: they have two spatial dimensions, like a classical image, and a third spectral dimension (energy or wavelength). In other words,  an energy spectrum  is associated to each spatial pixel. 

In recent decades, charge-coupled device (CCD) cameras have been the main spectro-imaging instruments on board X-ray telescopes such as XMM-Newton, Chandra, Swift, Suzaku, and eROSITA.
When the X-ray photon interacts with the semiconductor material,  an electric charge is generated proportional to the energy of the incident photon. 
The charge is then transferred row by row to a readout register.
For example, the XMM-Newton MOS camera can accommodate a large number (600$\times$600) of small  pixels \citep[40 $\mu$m,][]{Turner2001}.
The energy resolution is of the order 150 eV\footnote{\url{https://xmm-tools.cosmos.esa.int/external/xmm_user_support/documentation/uhb/epic_specres.html}} at 6 keV ($E \,/\Delta  E \sim 40$), which is close to the Fano limit, the physical lower bound of energy resolution due to the statistical fluctuations in the number of electron-hole pairs produced in the photon-material interaction.

To overcome this limit, a change of technology is required. Microcalorimeter devices are providing this revolutionary leap forward for X-ray spectro-imaging instruments. 
The fundamental principle of microcalorimeters is to measure the photon energy via the small increase in temperature caused by the conversion of this energy into heat in the sensor. To be sensitive to small heat increments and reach a spectral resolution on the order of a few electron volts, the device has to be cooled down to 50 mK and remain at very stable temperature.
It is important to note that, unlike gratings on board XMM-Newton or Chandra, microcalorimeters are non-dispersive spectrometers and can be used to study diffuse structure.

However, with the current available technology, only a limited number of pixels can be accommodated on the focal plane.
The XRISM/Resolve instrument is a microcalorimeter 
 array of 6 × 6 sensors \citep[pixel size of 800 $\mu$m,][]{hitomi_spie2014} with a {constant} spectral resolution over the entire bandpass of 5 eV FWHM \citep{XRISM_white_paper2020}.
While future X-ray telescopes combining microcalorimeter and high spatial resolution are planned  \citep[Athena/XIFU, see ][for more details]{Barret2018}, they are expected to launch in the late 2030s, and will still not reach the spatial resolution of the X-ray telescope Chandra.

In the meantime, mitigation plans to overcome this spatial-resolution limitation should be investigated to try to partially recover some of the spatial information. To that end, we explore in this study the performance of hyperspectral fusion to combine data sets with high spectral resolution and high spatial resolution using the example of fusion between XMM-Newton and XRISM/Resolve data.
The final goal of this reconstruction is to enhance the scientific return of XRISM/Resolve by retrieving maps of physical parameters (plasma temperature, velocity, abundances, etc.) at higher resolution with reduced cross-contamination between pixels (spectrospatial mixing).

Indeed, this mixing is an important problem when analyzing hyperspectral data. If an instrument's spatial resolution is larger than the coherence scale of a physical parameter, the reconstructed information will be an average of the small-scale features.
To take the example of a velocity map from the ejecta of a SNR, the small-scale redshifted and blueshifted ejecta features can mistakenly be measured at small or close-to-zero redshift values, as was demonstrated by the detailed analysis of Tycho's SNR in  \cite{Godinaud_2023}. By jointly deconvolving the data sets from two X-ray instruments, hyperspectral data fusion could solve this problem and recover higher-fidelity  scientific information at small scale.



\subsection{Hyperspectral fusion}
\label{sec:lit_review}
Given two hyperspectral (HS) images, the general task of hyperspectral fusion consists in obtaining a cube with the best spatial and spectral resolution of each data set. For instance, Fig. \ref{fig:Hitomi_XMM} shows HS images of the Perseus galaxy cluster taken by the X-ray telescopes XMM-Newton and Hitomi/SXS\footnote{XRISM/Resolve camera is identical to the SXS on board the short-lived Hitomi telescope launched in 2016.}. XMM-Newton has better spatial resolution, and Hitomi has better spectral resolution. The goal of fusing the two datasets would be to obtain a HS image with XMM-Newton's spatial resolution and Hitomi's spectral resolution. 

This problem has been explored in the literature for applications to satellite pictures of the earth, and has recently received some attention in astrophysics with the launch of the James Webb Space Telescope (JWST). However, X-ray astrophysics has its own specific challenges. X-ray data are dominated by Poisson noise, and both instruments produce HS images with many spectral channels, as opposed to one of the data sets being a multi-spectral image with a small number of spectral bands typically obtained with different filters. To use hyperspectral fusion for X-ray astrophysics, it is thus important to find the fusion method most appropriate for this case study. 

In this section, we provide an overview of the various aspects to take into account when approaching a problem of hyperspectral fusion in order to find the best way to solve the problem. For a detailed review of the literature, we refer the interested reader to  \cite{fusionreview}. 

The first aspect to consider is the forward model of the two instruments. This defines how a source image will be degraded in the process of image formation and recording, which depends on the systematic characteristics of the instrument, and the noise we expect to measure.

When considering systematic instrumental degradation, two dimensions may be taken into account: the spatial degradation, and the spectral degradation. Spatially, the degradation will be due to optical diffraction or other systematic blurs, and this information is encapsulated in the point spread function (PSF). Given a point source, the PSF returns its radiance distribution on the recorded image. Similarly, if a source emits a Dirac spectrum, the measured spectrum will be the instrument's spectral response function, and this will be affected by, for instance, the chemical and physical properties of the spectro-imager. Depending on the instrument, the PSF and spectral response may depend on the energy, on the spatial location of the pixel on the camera, or both. 

Then, for the stochastic noise, while most of the literature on fusion assumes additive Gaussian noise, it is important to realize this is not always appropriate. In X-ray astrophysics, photons have a low flux and are measured individually, and thus follow a Poisson distribution. 

\begin{figure*}[t!]
    \centering
    \includegraphics[width=\textwidth]{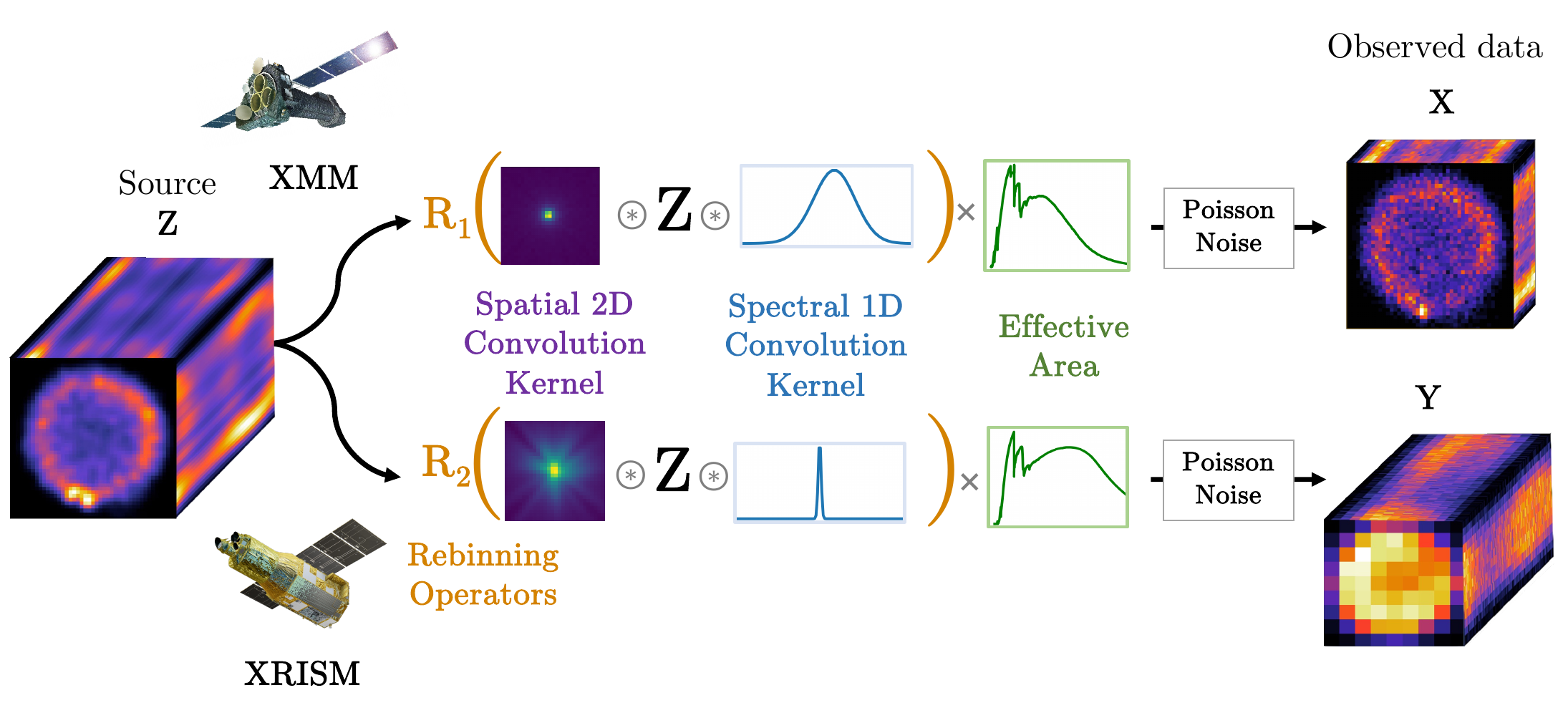}
    \caption{Simplified diagram of the forward model of two X-ray telescopes (here, XMM-Newton/EPIC and XRISM/Resolve). The problem of fusion consists in inverting this model and using the observed data X and Y to retrieve the deconvoluted cube Z. The mathematical details behind the operators and kernels are detailed in section \ref{sec:methods}. Not shown is the exposure time of each data set.}
    \label{fig:diagram_fusion}
\end{figure*}

Once the instrumental systematic degradation and the stochastic noise have been taken into account, if we take two HS images $X$ and $Y$ of sizes ($M\times N \times l$) and ($m \times n \times L$), the general problem of fusing two HS images can be written as: 

\begin{equation}
    \hat{Z}= \argmin_{Z} D (\mathcal{M}_X(Z)| X) + D(\mathcal{M}_Y(Z)| Y),
\end{equation}

where $\hat{Z}$ is the fusion result, $\mathcal{M}_X$ and $\mathcal{M}_Y$ are the model operators that take into account the systematic instrumental degradation (PSF, spectral response, etc.) and the rebinning needed to take account different voxel sizes, and $D$ is a well-chosen function that evaluates the difference between the model $\mathcal{M}_X(Z)$ and data $X$, and between model $\mathcal{M}_Y(Z)$ and data $Y$,
given the expected noise distribution. Thus, this problem ensures that $Z$ is a plausible source from which the first instrument can measure $X$ and the second can measure $Y$. In the case of X-ray astrophysics, the forward model is the one described in section \ref{sec:forwardmodel} and Fig. \ref{fig:diagram_fusion}.

However, this problem is often ill-posed \citep[in the sense of][]{Hadamard} and ill-conditioned, meaning that while there might be a unique solution, the noise will be amplified when inverting the model, corrupting the solution, and slight changes in the data would lead to drastic changes in the solution \citep{Bertero_2022}. One well-studied approach to solving ill-posed inverse problems is to regularize the problem, which restricts the space of acceptable solutions. This can be approached by considering the spatial features of $Z$, its spectral features, or both. To regularize in the spectral dimension, one classic method is to project the spectra of $Z$ onto a lower-dimensional subspace, as is done with source separation \citep[e.g.,][]{CNMF,prevost2022,pineau_2023}
or low-rank approximation \citep[e.g.,][]{HySure,tensor,guilloteau_2020}.
We can also add a regularization term $\varphi(Z)$ to the function we seek to minimize, which leads us to: 
 \begin{equation}
    \hat{Z}= \argmin_{Z} D (\mathcal{M}_X(Z)| X) + D(\mathcal{M}_Y(Z)| Y) + \varphi(Z).
    \label{eq:gen_problem}
\end{equation}

This $\varphi(Z)$ will promote certain spatial or spectral structures in the solution. Its choice must thus be appropriate for our data. For instance, in remote sensing (i.e., the study of satellite pictures of the earth)
, it may be useful to promote spatial piece-wise smoothness using a total variation term, as done by \cite{HySure}, but this would be ill-advised for astrophysical data, where such features are rarely observed. For our case of X-ray astrophysics, we consider three potential regularization approaches, which are described in section \ref{sec:regularization}. 

Finally, to solve the problem in equation \ref{eq:gen_problem}, we need to choose an appropriate optimization scheme. In some cases, there might be an analytical solution (as in the work of \cite{pineau_2023}  for JWST, when the spectral content is assumed to be a linear combination of known spectra), but most often an iterative algorithm must be constructed. The choice of optimizer will depend on the data size, the specific challenges presented by the forward model, and the chosen regularization terms. It may also be the case that constraints placed on the optimizer (such as insisting that the algorithm reaches convergence as quickly as possible) will lead to constraints on the choice of regularization and cost function.

In astrophysics, work has been done to tackle multi-spectral/HS fusion for the JWST \citep[see][]{guilloteau_spatreg_2022,pineau_2023} in order to fuse images taken by two of its instruments (NIRCam and NIRSpec), but to our knowledge, no algorithm has been implemented for HS/HS fusion with Poisson statistics, considering two instruments with their PSF and spectral responses, as is needed in X-ray astrophysics. 

\section{Methodology}
\label{sec:methods}

In this section, we present the methodology of our fusion algorithm for X-ray astrophysics. We present the general problem to be solved in section \ref{sec:forwardmodel}. Section \ref{sec:rebin} presents the rebinning operator needed when the data sets have different voxel sizes. Section \ref{sec:regularization} describes our choices of potential regularization terms. Section \ref{sec:algo} presents our algorithmic strategy to solve the inverse problem.

\subsection{Forward model}
\label{sec:forwardmodel}


Let $X$ be the hyperspectral data from the instrument with higher spatial resolution but lower spectral resolution (e.g., XMM-Newton/EPIC). Let 
$Y$ be the data from the other instrument, with lower spatial resolution and higher spectral resolution (e.g., XRISM/Resolve). When available, $\bar{{X}}$ and $\bar{{Y}}$ are the ground truths at the resolutions of the respective instruments. 

In this section, we consider the problem where both HS images are of the same size, that is,  $X$ and $Y$ $\in \mathbb{R}^{M\times N \times L}_+$, and therefore so does $Z$ (see section \ref{sec:rebin} for the case where $X$ and $Y$ have different sizes). Our forward model (seen in Fig. \ref{fig:diagram_fusion}) must include the spatial degradation of  each instrument, which is modeled by their PSFs, as well as the spectral degradation, which is modeled by their spectral responses. We must also account for the effective area of the instrument, and the exposure time. For the former, we assume an effective area with no vignetting; that is, the effective area does not vary spatially.\footnote{However, accounting for vignetting would be easy, since the effective area is simply multiplied to the model term by term. Thus whether we consider the effective area as a spatially uniform cube or a vignetted cube has no influence on the algorithmic complexity.} We use $A_X$ and $A_Y$ to denote the effective area cubes, which have the same shape as $X$ and $Y$. 
$t_X$ and $t_Y$ are scalars, the exposure times of each data set. The spectral responses of the  instruments are $\textbf{s}_X$ and $\textbf{s}_Y$, and their PSFs are  $B_X$ and  $B_Y$. 

We define $\mathcal{B}_X$ and $\mathcal{B}_Y$ as the operators which, given a cube $Z$, apply a 2D convolution on its spatial slices with the PSF $B_X$ or $B_Y$. Similarly, we use $\mathcal{S}_X$ and $\mathcal{S}_Y$ to denote the operators that convolve each spectra of $Z$ with $\textbf{s}_X$ or $\textbf{s}_Y$, respectively. Formally, if $Z_k$ is the 2D slice of $Z$ at the $k$th spectral channel, and $Z_{ij}$ is the spectra of $Z$ at pixel $(i,j)$: 

\begin{flalign}
    &\Big( \mathcal{B}_Y Z\Big)_k=B_Y \circledast Z_k ,\\
    &\Big( Z \mathcal{S}_X \Big)_{ij}=  Z_{ij} \circledast S_X,
\end{flalign}

where $\circledast$ represents the convolution operator (for simplicity of notation, this is written in the same way for 2D and 1D convolution). 

The aim is to reach, as closely as possible given the noisy data, a fused cube $Z$ that has the spatial resolution of $X$ and the spectral resolution of $Y$. The general problem is
\begin{equation}
    \hat{Z}=\argmin_{Z\geq 0} \mathcal{L}_P\Big(X \;| \;Z_X\Big)+\mathcal{L}_P\Big(Y \;| \;Z_Y\Big)
    + \varphi(Z),
    \label{eq:problem_norebin}
\end{equation}
where 
\begin{eqnarray}
    &Z_X=(t_XZ\odot A_X)\mathcal{S}_X \nonumber \\
    &Z_Y=\mathcal{B}_Y(t_YZ\odot A_Y),
    \label{eq:forwardmodel}
\end{eqnarray}
    
where $\odot$ is the term-by-term product, the data-fidelity metric $\mathcal{L}_P$ is the negative Poisson log-likelihood, and $\varphi(Z)$ is the regularization term. We impose $Z\geq 0,$ as there is no physical interpretation for a negative spectra. The likelihood terms promote faithfulness between our model and the data at comparable resolutions. 

It is worth noting that in order to fully invert the forward model, we would need to compare $X$ with $\mathcal{B}_X(t_XZ\odot A_X)\mathcal{S}_X$, and likewise for $Y$. However, our aim is not to retrieve a cube $Z$ with infinite resolution, but rather our intention is for $Z$ to have the spatial resolution of $X$, and the spectral resolution $Y$. The problem expressed in Equations \ref{eq:problem_norebin} and \ref{eq:forwardmodel} will promote such a solution, since no spatial degradation is applied to go from $Z$ to $Z_X$, and likewise spectrally for $Z_Y$ with $Y$.   

More explicitly, we obtain
\begin{flalign}
    \hat{Z}=&\argmin_{Z\geq 0}
    \sum_{i=1}^M\sum_{j=1}^N\sum_{k=1}^L \Bigg[
    \Big((t_XZ\odot A_X)\mathcal{S}_X\Big)_{ijk} \nonumber \\
    &- X_{ijk} ln\Big((t_XZ\odot A_X)\mathcal{S}_X\Big)_{ijk}
    \nonumber \\
    &+\Big(\mathcal{B}_Y(t_YZ\odot A_Y)\Big)_{ijk} - Y_{ijk} ln\Big(\mathcal{B}_Y(t_YZ\odot A_Y)\Big)_{ijk}\Bigg]
    + \varphi(Z),
    \label{eq:Poisson_problem_norebin}
\end{flalign}
where the shape of $Z$, $X$, $Y$, are all $(M,N,L)$.

\subsection{Rebinning operator}
\label{sec:rebin}
In most cases, two different hyperspectral instruments will not produce data of the same voxel size. Rather, we will have $X \in \mathbb{R}_+^{M\times N \times l}$ and  $Y \in \mathbb{R}_+^{m\times n \times L}$, where $L>l$, $M>m$, and $N>n$. In other words, $X$ has more pixels than $Y$, and $Y$ has more spectral channels than $X$. Our result $Z$ should be of size $(M\times N \times L)$. In order to compare $Z$ to the data $X$ and $Y$ in the data fidelity terms of equation \ref{eq:problem_norebin}, we thus need to change the size of $Z$ to that of $X$ and $Y,$ respectively. This is done by applying a rebinning operator. 

For the sake of simplicity, and to keep this operator differentiable, we assume that $M$ is a multiple of $m$, $N$ is a multiple of $n$, and $L$ is a multiple of $l$. Let us denote the multiplying factors $(\nu_M,\nu_N,\nu_L)$, such that $M=\nu_M m$, and likewise for the other dimensions. We need three rebinning operators: $\mathcal{R}_{M}$, $\mathcal{R}_{N}$, and $\mathcal{R}_{L}$. 

In this case, the rebinning operator is that which 
sums together pixels in nearby groups of $\nu$. 

For instance, for $\mathcal{R}_{M}$, a single voxel $(s,j,k)$ of the rebinned tensor is
\begin{equation}
    \big(\mathcal{R}_M\textbf{Z}\big)(s,j,k)=\sum_{p=s*\nu_M}^{p=(s+1)*\nu_M-1}  \textbf{Z}(p,j,k).
\end{equation}
Such an operator can be neatly written as a matrix product between each slice of the tensor $Z$ and a block diagonal matrix filled only with zeroes and ones. For simplicity, we use $\mathcal{R}_{MN}$ to denote the spatial rebinning operator (which applies $\mathcal{R}_M$ on the lines, and $\mathcal{R}_N$ on the columns).

Now, to include rebinning in our optimization problem posed in equation \ref{eq:problem_norebin}, $Z_X$ and $Z_Y$ become
\begin{flalign}
    &Z_X=(t_XZ\odot A_X)\mathcal{S}_X\mathcal{R}_L \nonumber \\
    &Z_Y=\mathcal{R}_{MN}\mathcal{B}_Y(t_YZ\odot A_Y),
    \label{eq:forwardmodel_rebin}
\end{flalign}

and plugging this into the Poisson likelihood, we obtain 
\begin{flalign}
    \hat{Z}=&\argmin_{Z\geq 0}
    \sum_{i=1}^M\sum_{j=1}^N\sum_{k=1}^L \Bigg[
    \Big((t_XZ\odot A_X)\mathcal{S}_X\mathcal{R}_L\Big)
    _{ijk} \nonumber \\
    &- X_{ijk} ln\Big((t_XZ\odot A_X)\mathcal{S}_X\mathcal{R}_L\Big)_{ijk}
    +\Big(\mathcal{R}_{MN}\mathcal{B}_Y(t_YZ\odot A_Y)\Big)_{ijk} \nonumber \\
    &- Y_{ijk} ln\Big(\mathcal{R}_{MN}\mathcal{B}_Y(t_YZ\odot A_Y)\Big)_{ijk}\Bigg]
    + \varphi(Z).
    \label{eq:Poisson_problem_rebin}
\end{flalign}

\subsection{Regularization}
\label{sec:regularization}

In X-ray astrophysics, low signal-to-noise ratio presents a major problem, as HS images are dominated by Poisson noise. Thus, we are looking for regularization terms that will be efficient denoisers for Poisson statistics, and that will be appropriate for astrophysical images (often displaying diffuse and/or isotropic features) and X-ray spectra (which can present strong spectral variability). 

Our first consideration was sparsity under a 2D-1D starlet transform. Then, inspired by the work of \cite{guilloteau_spatreg_2022} regarding JWST, we implemented a low-rank approximation with Sobolev spatial regularization. Finally, we wished to test a compromise between the two, with a low-rank approximation and a 2D spatial wavelet sparsity regularization. The following sections describe these three options. 

\subsubsection{2D-1D wavelet sparsity}
A classic regularization method is to choose a term that promotes the sparsity of $Z$ under a well-chosen orthogonal transform — that is to say, that in some domain, $Z$ can be represented by only a small number of non-zero coefficients. 
Wavelets are useful for capturing multi-scale features. In particular, the isotropic undecimated wavelet transform, dubbed the starlet transform, is appropriate for astrophysical data due to its isotropic structure \citep{starlet}. The data are convolved by $B_3$-spline filters, which produce a set of increasingly coarse resolution versions of the data, and the $S$ wavelet scales are defined as the difference between two subsequent resolutions. The coarsest resolution is dubbed the coarse scale, and no assumption of sparsity is placed upon it. This allows the regularization to represent data with a smooth continuum (the coarse scale) and sharp features (the non-zero elements of the fine scales).

More specifically, we chose the 2D-1D starlet transform, which we define as applying the 2D starlet transform channel-wise, then the 1D starlet transform pixel-wise. Thus, in this case, $\varphi(Z)$ is the $l_1$ norm \citep[the minimization of which promotes sparsity; see][]{sparsity} of the 2D-1D starlet transform coefficients of $Z$. For $S$ wavelet scales, we define the 2D-1D starlet transform as $\mathcal{W}_S^{2D1D}$, and the $l_1$ norm regularization term is
\begin{flalign}
    \varphi(Z)&=\Big|\Big| \mathcal{W}_S(Z) \Big|\Big|_1 \nonumber \\
    &=\mu \sum_{i=1}^M\sum_{j=1}^N\sum_{k=1}^L\sum_{s=1}^S|\mathcal{W}_S(Z)^{2D1D}_{i,j,k,s}|,
\end{flalign}
where $\mu$ is a hyperparameter to control the intensity of regularization.

\subsubsection{Low rank and Sobolev}
This regularization option was inspired by work regarding JWST carried out by  \cite{guilloteau_2020}, though we designed an implementation adapted for HS-HS fusion and the specifics of X-ray astronomy. The spectral regularization is a low-rank approximation, which corresponds to decomposing $Z= W V,$ where $V\in \mathbb{R}^{L\times r}$ spans the subspace of rank $r$ onto which $Z$ is projected, and $W \in \mathbb{R}^{r \times M \times N}$  is the tensor of representation coefficients in that subspace. $V$ is found by performing principal component analysis on $Y$ (the hyperspectral image with the most spectral information) and keeping only the highest singular values. Therefore, now $Z_X$ and $Z_Y$ from Equation \ref{eq:forwardmodel_rebin} become variables of $W$, where

\begin{flalign}
    &Z_X(W)=(t_X WV\odot A_X)\mathcal{S}_X\mathcal{R}_L \nonumber \\
    &Z_Y(W)=\mathcal{R}_{MN}\mathcal{B}_Y(t_Y WV \odot A_Y),
    \label{eq:forwardmodel_lowrank}
\end{flalign}

and these can be plugged into Equation \ref{eq:Poisson_problem_rebin}, now seeking to minimize $W$. 
For spatial regularization, the Sobolev regularization \citep{Sobolev} implies adding the following term to the cost function:
\begin{equation}
    \varphi(W)=\mu ||\mathcal{D}(W)||^2_F,
\end{equation}
where $\mathcal{D}$ is a first-order 2D finite difference operator. This term will act to minimize the differences between neighboring pixels on the representation coefficients of the subspace spanned by $V$.

Both the Sobolev regularization and the low-rankedness will result in a faster algorithm, due to the dimension reduction and simple expression of the $\mathcal{D}$ operator. However, Sobolev does not capture multi-scale spatial features. Further, as $V$ is extracted from $Y$, it may reproduce features found in $Y$ too strongly. This is acceptable (and even desirable) in $MS/HS$ fusion, where the MS image does not bring a significant amount of spectral information, but in $HS/HS$ fusion, $X$ contains spectral information, in particular because it displays much less spatial mixing than $Y$. The assumption that all spectra in $Z$ can be decomposed as linear combinations of $Y$ is not necessarily true, and forcing a low-rank approximation in this case will bring on spectral distortions. However, if this assumption is valid, low-rankedness will be an undeniable asset.

\subsubsection{Low rank and 2D wavelet sparsity}
As Sobolev regularization does not capture multi-scale information in the same way wavelets do, we considered the option to keep a low-rank approximation as described in the previous section, and then to use a starlet sparsity term for the spatial regularization: 
\begin{equation}
    \varphi (W)=\mu ||\mathcal{W}_S^{2D}(W)||_1,
\end{equation}
where the operator $\mathcal{W}_S^{2D}$ applies the 2D starlet transform to every slice of $W$.

\subsection{Algorithm description}
\label{sec:algo}

\subsubsection{Proximal gradient descent}
If we consider the problem posed in Equations \ref{eq:Poisson_problem_norebin} or \ref{eq:Poisson_problem_rebin}, and replace the regularization term by one of the three options we consider above, we note that the cost function is convex, and can be separated into two parts, one differentiable and one non-differentiable. 
In such cases, proximal gradient methods have been shown to be very effective \citep{prox}. 

The proximal gradient descent is an iterative method where, for each iteration, we first undergo a gradient descent step on the differentiable part of the cost function before applying a proximal operator (\textbf{prox}), the purpose of which is to find a solution close to the result obtained by gradient descent while minimizing the function's non-differentiable part. Formally, if the result obtained by the gradient descent step is $x$, and the non-differentiable part of the function is $\varphi$, the operator is:
\begin{equation}
    \textbf{prox}_{\varphi}(x)=\argmin_y(\varphi(y)+\frac{1}{2}||y-x||_2^2).
\end{equation}
For the $l_1$ norm, the proximal operator is a soft thresholding function:
\begin{equation}
   \textbf{prox}_{l_1,\mu}(x)=
    \begin{cases}
     0 & \text{if } |x| < \mu\\
     x - \mu \cdot sign(x) & \text{if } |x| \geq \mu \\
    \end{cases},  
    \label{eq:proxl1}
\end{equation} 
where $\mu$ is the parameter that controls the strength of regularization. For the non-negativity constraint, the \textbf{prox} is simply the operator that forces all negative values to 0. 

The gradient of the differentiable part in Equation \ref{eq:problem_norebin} is 
\begin{flalign}
    \nabla_{Z}\mathcal{L}_{P}&=\nabla_{Z}\Big(\mathcal{L}_P(X \;| \;Z_X)\Big)+\nabla_{Z}\Big(\mathcal{L}_P(Y \;| \;Z_Y)\Big) \nonumber \\
    &=\frac{\partial(Z_X)}{\partial{Z}}\nabla_{Z_X}\Big(\mathcal{L}_P(X \;| \;Z_X)\Big)+\frac{\partial(Z_Y)}{\partial{Z}}\nabla_{Z_Y}\Big(\mathcal{L}_P(Y \;| \;Z_Y)\Big).
    \label{eq:grad_poisson}
\end{flalign}

In the case with rebinning, we have 
\begin{flalign}
&\frac{\partial(Z_X)}{\partial{Z}}=t_X A_X\mathcal{R}_L^{\dagger} \mathcal{S}_X^{\dagger}, \nonumber \\
&\frac{\partial(Z_Y)}{\partial{Z}}=t_Y A_Y \mathcal{B}_Y^{\dagger}\mathcal{R}_{MN}^{\dagger},
\end{flalign}
where $\dagger$ denotes the conjugate transpose of the operators, in the sense that, if we were to vectorize the spatial dimensions of our data and write the operator $\mathcal{S}$ as a product with the 2D matrice $S$,  $\mathcal{S}^{\dagger}$ would apply the product with matrix $S^{\dagger}$. In the case with a low-rank approximation, where the gradient descent is done on $W=ZV^T$:
\begin{flalign}
&\frac{\partial(Z_X)}{\partial{W}}=V t_X A_X \mathcal{R}_L^{\dagger} \mathcal{S}_X^{\dagger}, \nonumber \\
&\frac{\partial(Z_Y)}{\partial{W}}=V t_Y A_Y \mathcal{B}_Y^{\dagger}\mathcal{R}_{MN}^{\dagger}.
\end{flalign}

In the case of Sobolev regularization, $\varphi_{Sobolev}(W)=\mu ||\mathcal{D}(W)||^2_F$ is differentiable, and so a term is added to the gradient: 
\begin{equation}
    \varphi_{Sobolev}'(W)=2\mu \mathcal{D}^T\mathcal{D}(W).
    \label{eq:sobolev_grad}
\end{equation}
As is the case for the other operators, $\mathcal{D}$ can be interpreted as a product with the 2D matrix $D$, since the operator $\mathcal{D}$ is a convolution with a kernel of $[1,-1]$ on the lines followed by a convolution with a kernel of $[1,-1]^T$ on the columns. $\mathcal{D}^T$ applies the product with $D^T$.

In the code, the operators $\mathcal{R}$ and $\mathcal{D}$ are written as matrix products. The operators $\mathcal{B}_Y$ and $\mathcal{S}_X$, however, can be dealt with more efficiently by taking advantage of the diagonalization of the convolution operation in the Fourier domain. 
To do so, we pad the kernels and $Z$ with zeros so that they are the same size (dimension of $Z$ + dimension of the kernel in question). We then take their Fourier transform, and convolution becomes a Hadamar product in the Fourier space. The Fourier transform of $Z$ will need to be calculated at each iteration, but the transform of the kernels need only be calculated once. 

Putting all this together, we obtain Algorithm \ref{alg:fusion}, which we have dubbed \texttt{HIFReD}: Hyperspectral Image Fusion with Regularized Deconvolution. For simplicity, rebinning is not explicit in the pseudo-code.\\ \\  
\begin{algorithm}[h!]
\setstretch{1.1}
    \caption{\texttt{HIFReD}: Hyperspectral Image Fusion with Regularized Deconvolution}
    \DontPrintSemicolon
    \textbf{Input:} data ($X$, $Y$), instrument response kernels ($S_X$, $B_Y$), regularization function ($\varphi$), Low Rank (True/False), rank $r$, regularization parameter $\mu$, number of wavelet scales $S_{2D},S_{1D}$, maximum number of iterations $t_{max}$.
    
    \texttt{PWS}: Proximal Wavelet Step, defined in Alg. \ref{alg:Wprox}.
    
    \textbf{Output:} $Z$
    
    \textbf{\textit{Initialization}}\linebreak
    Calculate $\mathcal{F}^{1D}(S_X), \mathcal{F}^{2D}(B_Y)$. \linebreak
    $Z \gets X$\linebreak
    \eIf{Low Rank}{
    $\hat{V} \gets$ eigenvectors of $(Y-\bar{Y})(Y-\bar{Y})^T$
    
    $V \gets \hat{V}[0:r,.]$
    }{
    $V\gets 1$ 
    }
    $t \leftarrow 0$
    
    \While  {$t < t_{max}$}{
        $W \gets V^T Z$\linebreak
        $\tilde{W_t}=\mathcal{F}^{1D}(\mathcal{F}^{2D}(W_t))$. \linebreak
        \textbf{\textit{—Gradient Step—}}\linebreak
        \eIf {$\varphi(W)=||\mathcal{D}(W)||_F^2$}{
            $\tilde{W_{t+1}}\gets \tilde{W_{t}} - \alpha\Big[\nabla_{\tilde{W}}\mathcal{L}_P(\tilde{W_t}) -\mu \varphi_{Sobolev}'(W_t)\Big]$ \quad \quad Eq. \ref{eq:grad_poisson}, \ref{eq:sobolev_grad}.
            }{
            $\tilde{W_{t+1}}\gets \tilde{W_{t}} - \alpha\Big[\nabla_{\tilde{W}}\mathcal{L}_P(\tilde{W_t})\Big]\quad\quad $Eq. \ref{eq:grad_poisson}.
            }
            $W_{t+1} \gets \mathcal{IF}^{1D}(\mathcal{IF}^{2D}(\tilde{W_{t+1}}))$ \linebreak
        \textbf{\textit{—Proximal Step—}}\linebreak
        \If {$\varphi(W)=||\mathcal{W}_S^{2D}(W)||_1$}{
            $W_{t+1} \gets \texttt{PWS}\Big[W_{t+1},\mathcal{W}^{2D}_{\mu, S2D}\Big]$ 
            }
        \If {$\varphi(W)=||\mathcal{W}_S^{2D-1D}(W)||_1$}{
            $W_{t+1} \gets \texttt{PWS}\Big[W_{t+1},\mathcal{W}^{2D-1D}_{\mu, (S1D, S2D)}\Big]$ 
            }
        $Z_{t+1} \gets W_{t+1} V$
        
        $Z_{t+1} \gets Z_{t+1} \odot \delta (Z_{t+1} \geq 0)$
        
        $t \gets t+1$
    }
    \Return $Z_{t}$
\label{alg:fusion}
\end{algorithm}

\begin{algorithm}[h!]
\setstretch{1.1}
\caption{\texttt{PWS}: Proximal Wavelet Step}\label{alg:Wprox}
    \textbf{input} cube $Z$,  wavelet transform $\mathcal{W}$, regularization parameter $\mu$, number of wavelet scales $S$\linebreak
    $\{w_{c}, w_{1},...,w_{S}\} \leftarrow \mathcal{W}(Z)$ \linebreak
    $\rho \gets  \mu \, \textbf{mad} (w_{1,1})$ \quad (median absolute deviation) \linebreak
    \For{ $i$ in $\{1,...,S\}$}{
            $w_{i} \leftarrow \textbf{prox}_{l1}(w_{i},\rho)$ \quad Eq. \ref{eq:proxl1}
        }
        $\hat{Z}\leftarrow w_{c,c} + \sum_{i=1}^{S} w_i$ \linebreak
    Return( $\hat{Z}$ )
\end{algorithm}
\subsubsection{Hyperparameters}

For the choice of hyperparameters (namely the regularizing factor $\mu$, the gradient step $\alpha$, and the rank $r$), we need to take a heuristic approach. Indeed, the Poisson likelihood is not globally Lipschitz-continuous, and so we cannot analytically calculate an optimal gradient step, and there is no a priori best choice for the regularization parameters either. Further, we find that one set of hyperparameters that works best for one toy model is not necessarily ideal for another model. The photon count, level of spectral variability, shape of the effective area, and more are all factors that will impact the choice of best parameters. It would be very costly to compute every single combination of hyperparameters to find the best suited for every single model, but below, we describe some guidelines to help pick suitable parameters, if not the exact optimal value.

The gradient step should be taken as high as possible for a faster rate of convergence, but if it is taken too high, the likelihood will diverge dramatically. This is easily diagnosed, and is seen as huge jumps in the cost function, when it should rather decrease monotonously. This usually happens in early iterations, and so we checked the value of the likelihood periodically at the start (first 1000 iterations, which takes less than an hour), and then stopped if it became pathological. If it this was the case, we then lowered the gradient step and tried again with that parameter. As the Poisson likelihood is convex, there is no risk of falling into a local minimum, provided this step is well-chosen as described.

For the $\mu$ parameter of the Sobolev regularization, for the Gaussian toy model, we tested values of between 0.01 and 0.1, and found the best value to be 0.03. This value turned out to be well suited for the other toy models as well. An overly high value of this parameter is detectable as an over-smoothing of spatial features, while an excessively low value will be recognized in the presence of visible noise on spatial slices. 

For the rank $r$, we tested values of between 5 and 50, and selected 10. The rank has a similar impact to $\mu$, but on the spectral features. A lower rank might over-smooth and lead to a biased result, while a higher rank might not be enough of a denoiser. On top of that, the rank will impact the algorithm's computational complexity, with a lower rank implying fewer dimensions and thus a faster algorithm. 

For the wavelet threshold parameter $\mu$, this is usually taken to be between 0.5 and 3, which in the case of Gaussian noise is comparable to a 0.5$\sigma$-3 $\sigma$ interval. There is no such easy interpretation for Poisson noise, but even so, like with the Sobolev, lower $\mu$ will allow more variability and less bias, while higher $\mu$ will lead to more smoothing, and thus potentially more bias. In particular, we find that in dim voxels (close to zero, and dominated by noise), higher values of $\mu$ (close to 3) are effective in smoothing out noise, while in brighter voxels, those high $\mu$ will over-smooth spectral features. We find the opposite to be true with lower values of $\mu$. As a compromise, and to account for the high variabilities of flux between voxels,  
it is advantageous to apply a varying $\mu$, using a sigmoid function going from $0.5$ to 3. Let the coarsest scale of $Z$ in the wavelet domain be $C_Z$. Further, let $\tilde{C}=|C_Z|-\bar{|C_Z|}$. In this case,
\begin{equation}
    \mu=\frac{(3-0.5)}{1 + e^{\tilde{C}}}+0.5,
\end{equation}
which will make $\mu(i,j,k)$ close to 1.25 when $|C_Z|$ is close to $\bar{|C_Z|}$, and go towards 0.5 for small values of $|C_Z|$, and towards 3 for high values of $|C_Z|$. This simple scheme is however ineffective in cases where the toy model is not very sparse in the wavelet domain, as is the case for the realistic model. In this case, it is better to pick a low, constant value of $\mu$, such as 0.5. Further, one needs to consider the number of wavelet scales defined in the starlet transform, be it the 2D or the 2D-1D transform. As the coarse scale is left untouched, the number of fine scales will determine the maximum size of the features that can be thresholded, as for each scale the size of the filter increases by a factor of 2 \citep{starckbook}. The number of 2D wavelet scales was chosen to be $2$ for all models, and the number of 1D scales was selected to be $3$ for the Gaussian model, and $1$ for the realistic model, for which the spectral details are much finer due to the finer spectral response.

The last hyperparameter to consider is $t_{max}$, the maximum number of iterations. The user should check the cost function curve at the end to see whether optimization has converged. If not, they may begin another batch of iterations, using the last result as first guess. An automatic stopping criterion could easily be implemented, with iterations stopping at iteration $t$ once $\Big| \frac{\mathcal{L}_P(Z_t)-\mathcal{L}_P(Z_{t+1})}{\mathcal{L}_P(Z_t)}\Big| > \epsilon$ for a small $\epsilon$, but the results in this article were obtained with a manually set $t_{max}$. 

\section{Results on simulated data}
\label{sec:results}

In this section, we present the results of our method on toy models derived from hydrodynamics simulations of a SNR. Section \ref{sec:toymodel} presents our three considered toy models: the Gaussian model, the Gaussian model with rebinning, and the realistic model with rebinning. Section \ref{sec:results_plots} then presents our results for each using the three regularization schemes described in section \ref{sec:regularization}. 
\begin{figure}[h!]
    \centering
    \includegraphics[width=\linewidth]{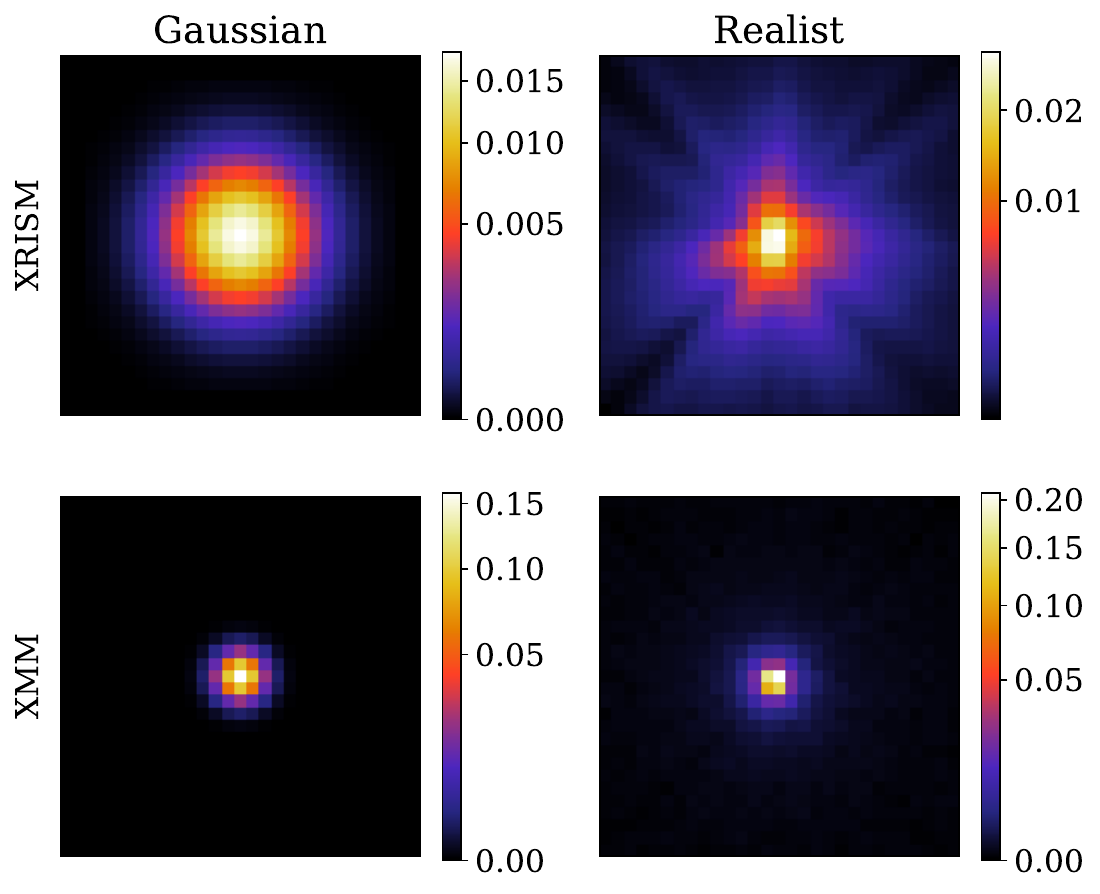}
    \caption{Comparison of the four PSFs used to generate the toy models. On the left, the Gaussian approximation of a XRISM-like and XMM-Newton-like instrument. On the right, the actual PSFs of XRISM/Resolve and XMM-Newton/EPIC. All PSFs are shown on images of size 4'50"$\times$4'50". Colors are displayed in square-root scale.}
    \label{fig:PSF_comp}
\end{figure}

\begin{figure}[h!]
    \centering
    \includegraphics[width=\linewidth]{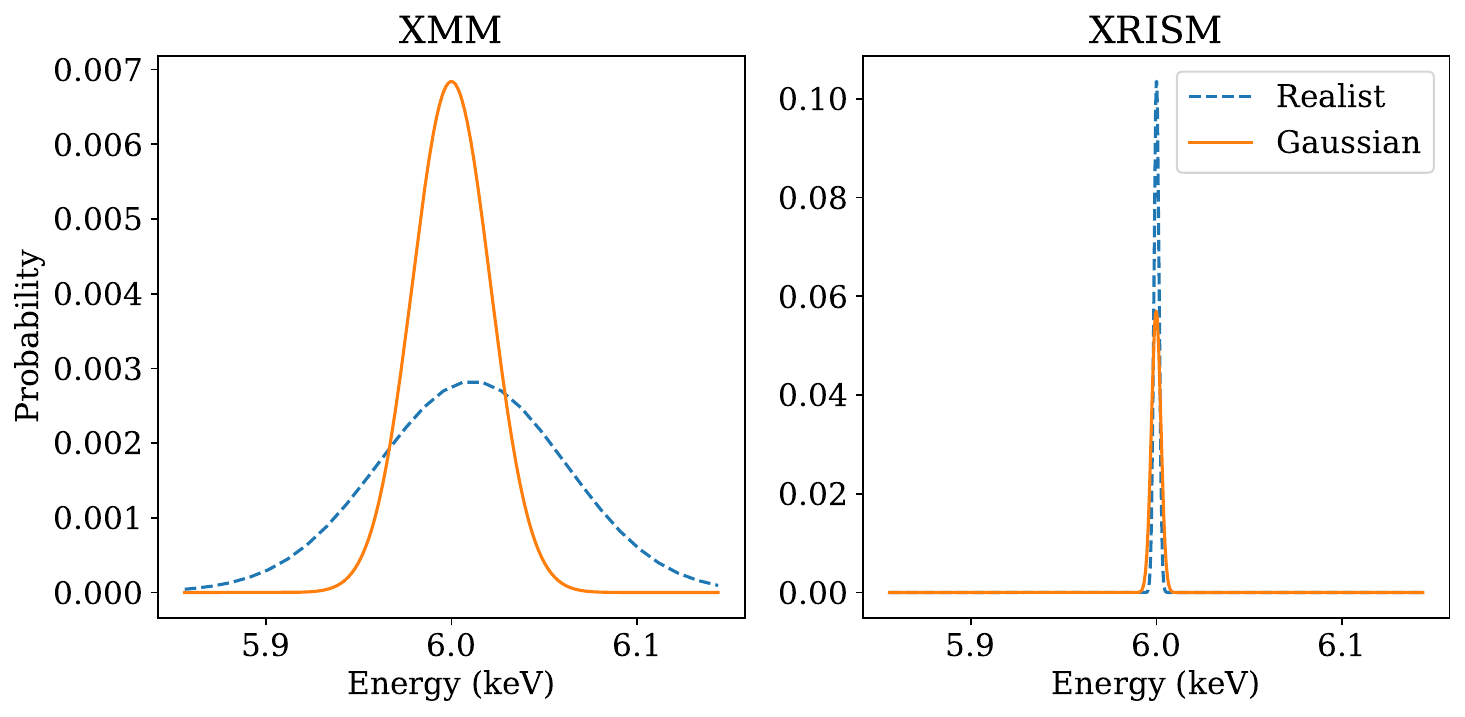}
    \caption{Comparison of the four spectral responses used to generate the toy models. The responses for the XMM-Newton-like instrument are shown on the right, while those for the XRISM-like are shown on the left. The full orange line is for the Gaussian approximations, and the dashed blue line is for the realistic instrument responses.}
    \label{fig:RMF_comp}
\end{figure}
\subsection{Toy models}
\label{sec:toymodel}
\begin{figure*}
    \centering
    \includegraphics[width=\linewidth]{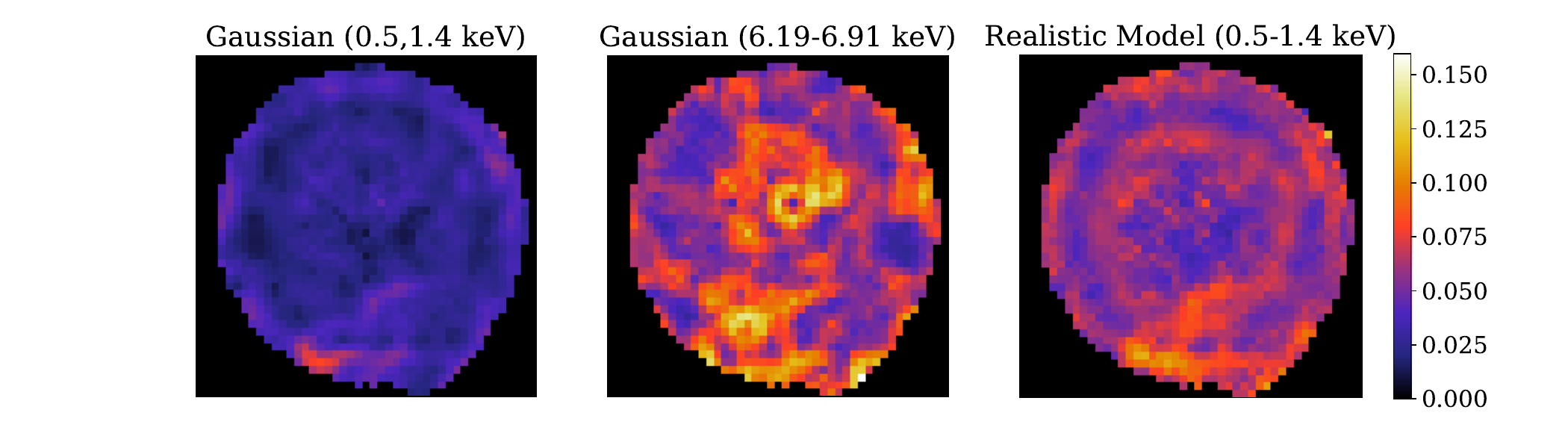}
    \caption{Average angular distance between neighboring pixels on a window of neighboring pixels, given by Equation \ref{eq:angular_distance_neighbourhood}. Higher values indicate a location of high spectral variability between pixels.}
    \label{fig:Angular_distance_TM}
\end{figure*}
The toy models we constructed are intended to resemble the case study of SNRs, such as Cassiopeia A. To this end, we used the 3D hydrodynamics simulations of \cite{Orlando_2016}. For each voxel of the 3D simulation, a synthetic X-ray spectrum was generated using the non-equilibrium ionization Xspec model NEI\footnote{\url{https://heasarc.gsfc.nasa.gov/docs/xanadu/xspec/manual/node200.html}}.
In the numerical simulation dataset, the temperature kT ranges from 0.2 keV to 3 keV, the ionization timescale $log(\tau)$ from 8 to 11.5 cm$^{-3}$ s, and the velocity along the line of sight $V_{\rm z}$ from -7000 to 7000 km s$^{-1}$. The absorption along the line of sight was fixed to $n_H = 0.5 \times 10^{22}$cm$^{-2}$.
Then the hypercube (X, Y, Z, E) is projected along the Z axis to emulate a synthetic X-ray observation of a SNR. 
Due to the projection effect, the spectrum for each pixel can be quite complex, and can contain, for example, double-peaked spectral features associated with the red- and blueshifted half spheres of the SNR or different temperatures and ionization timescales.

\def\arraystretch{1.5}
\begin{table*}[b!]
    \centering
    \caption{Characteristics of the simulated data sets.}
    \begin{tabular}{|>{\centering\arraybackslash}p{0.1\linewidth}|>{\centering\arraybackslash}p{0.1\linewidth}|>{\centering\arraybackslash}p{0.1\linewidth}|>{\centering\arraybackslash}p{0.1\linewidth}|>{\centering\arraybackslash}p{0.1\linewidth}|>{\centering\arraybackslash}p{0.1\linewidth}|>{\centering\arraybackslash}p{0.1\linewidth}|} \hline 
         Model&  \multicolumn{2}{c|}{\textbf{Gaussian} }&  \multicolumn{2}{c|}{\textbf{Gaussian (rebinning)}}&  \multicolumn{2}{c|}{\textbf{Realistic (rebinning)}}\\ \hline 
         Data set&  X&  Y&  X&  Y&  X& Y\\ \hline 
         Number of pixels &  \multicolumn{2}{c|}{45$\times$45 }&  45$\times$45&  15$\times$15& 45$\times$45&  15$\times$15 \\ \hline 
         Number of spectral channels&  \multicolumn{2}{c|}{a) 2500 - b) 2000 }& 60&  2400&  60 & 2400\\ \hline 
         Energy range&  \multicolumn{2}{c|}{a) 0.5-1.4 keV - b) 6.2-6.9 keV}&  \multicolumn{2}{c|}{0.5-1.4 keV} &  \multicolumn{2}{c|}{0.5-1.4 keV}\\ \hline 
         PSF std&  10" &  30" &  10" &  30" &  11"& 33" \\ \hline 
         Spectral response std&  21 eV&  2.5 eV&  21 eV&  2.5 eV&  51 eV& 1.4 eV\\ \hline
         Photon count&  a) $2.4\times 10^7$ b) $3.2\times 10^7$ &  a) $2.3\times 10^7$ b) $3.1\times 10^7$& $2.7\times 10^6$ & $2.5 \times 10^6$ &  $4.5 \times 10^{7}$&  $3.0 \times 10^{7}$\\ \hline
    \end{tabular}
    \tablefoot{The spectral and spatial responses are given in standard deviation (std).}
    \label{tab:comp_models}
\end{table*}

We then wished to consider different levels of complexity for our algorithm benchmarking. In particular, we considered three characteristics that can make the problem more complex: the intensity of spectral variability, the inclusion of a rebinning operator (X and Y have different voxel size), and the ratio between the spectral and spatial responses of the two data sets.

The first toy model we considered is dubbed the Gaussian model. The PSFs and spectral responses of each instrument are assumed to be Gaussian, whose parameters are summarized in Table \ref{tab:comp_models}, and shown in Fig. \ref{fig:PSF_comp} and \ref{fig:RMF_comp}. Further, we assumed a flat effective area, and that $X$ has the same size as $Y$. With these parameters, we generated one set of $X$ and $Y$  between 0.5 and 1.4 keV, and one between 6.2 and 6.9 keV (around the iron emission lines). The dataset around the iron lines displays more spectral variability, as is demonstrated more formally in Fig. \ref{fig:Angular_distance_TM}, which shows the average angular distance $\theta$ between neighboring pixels for each model. For a given pixel's spectra $Z_k$, it is calculated as 
\begin{equation}
\theta \big(Z_k\big)= \sum_{Z_{k'}\in \mathcal{N}} \arccos \Big( \frac{<Z_k,Z_{k'}>}{\sqrt(||Z_k||_2^2*||Z_{k'}||_2^2)}\Big)/\text{card} (\mathcal{N}),
    \label{eq:angular_distance_neighbourhood}
\end{equation}

where $\mathcal{N}$ is the set of pixels neighboring $Z_k$, that is, the eight closest pixels to $Z_k$, and $\text{card} (\mathcal{N})$ is the number of elements in that set. This angular distance provides information about the spectral variations across the remnant. As we see, the Gaussian model around the iron line has the strongest variations. 

Then, we considered the same Gaussian model (same PSF/spectral response, same flat effective area) between 0.5 and 1.4 keV, but this time with $X$ and $Y$ having different voxel sizes. $X$ has the dimensions ($45\times 45 \times 60$), with square pixels of 10" in width and energy channels of 14 eV in width, and $Y$ has the dimensions ($15\times 15 \times 2400$), with square pixels of 30" in width and energy channels of 0.36 eV in width. This allows us to see the impact of the rebinning operator on our algorithm.

Finally, we set out to obtain a more realistic model, which would mimic the case where we try to fuse data from XMM-Newton/EPIC and XRISM/Resolve. We chose PSFs and spectral responses meant to be as close as possible to those of the instruments in question (see Fig. \ref{fig:PSF_comp} and \ref{fig:RMF_comp}), and likewise for the effective areas. 
The energy range was (0.5-1.4 keV), and we chose the same dimensions and voxel size as the Gaussian model with rebinning. 

\begin{figure*}[h!]
    \centering
    \begin{subfigure}{0.48\textwidth}
      \centering
    \includegraphics[width=\textwidth]{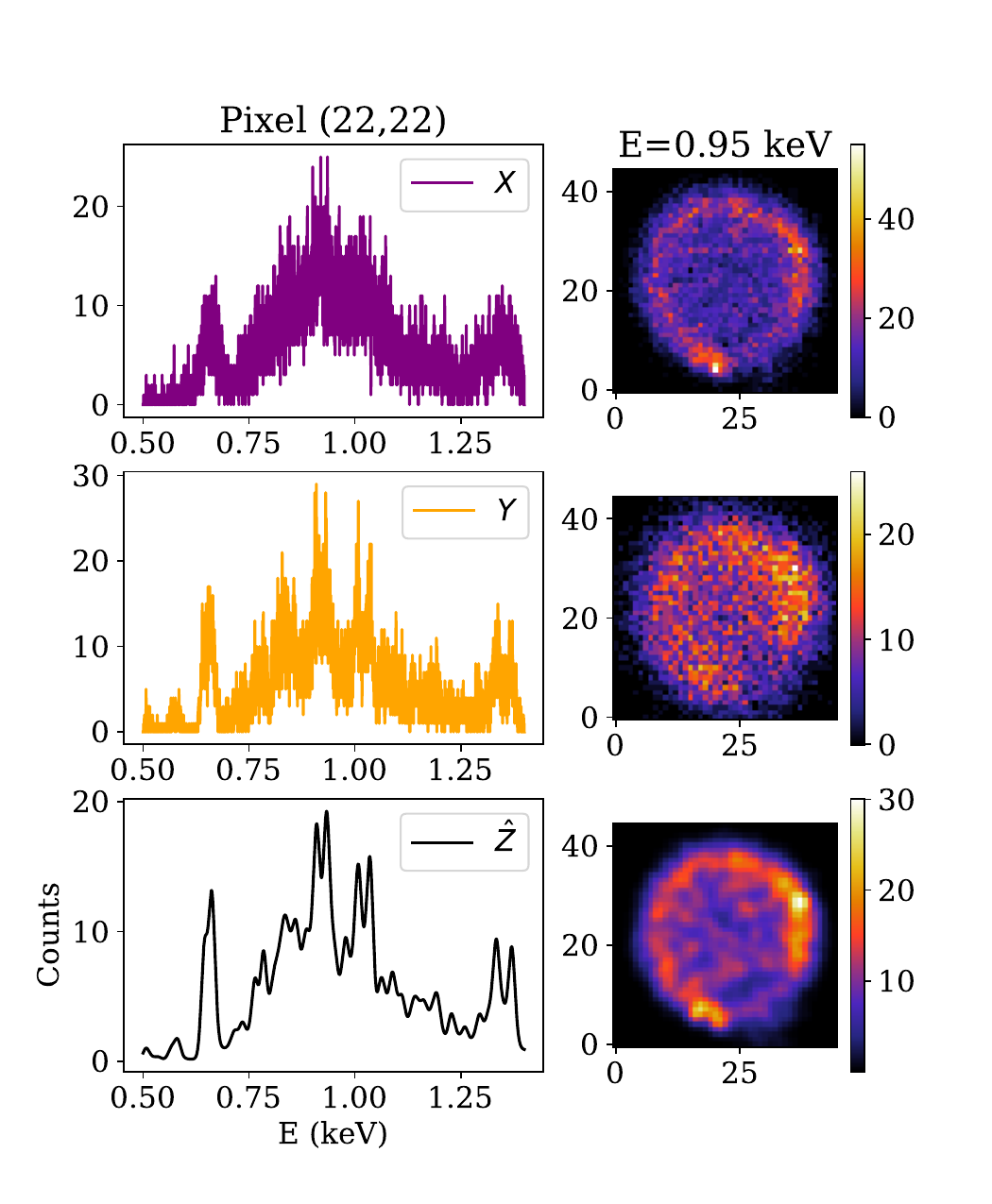}
    \caption{}
\end{subfigure}
    \begin{subfigure}{0.48\textwidth}
      \centering
    \includegraphics[width=\textwidth]{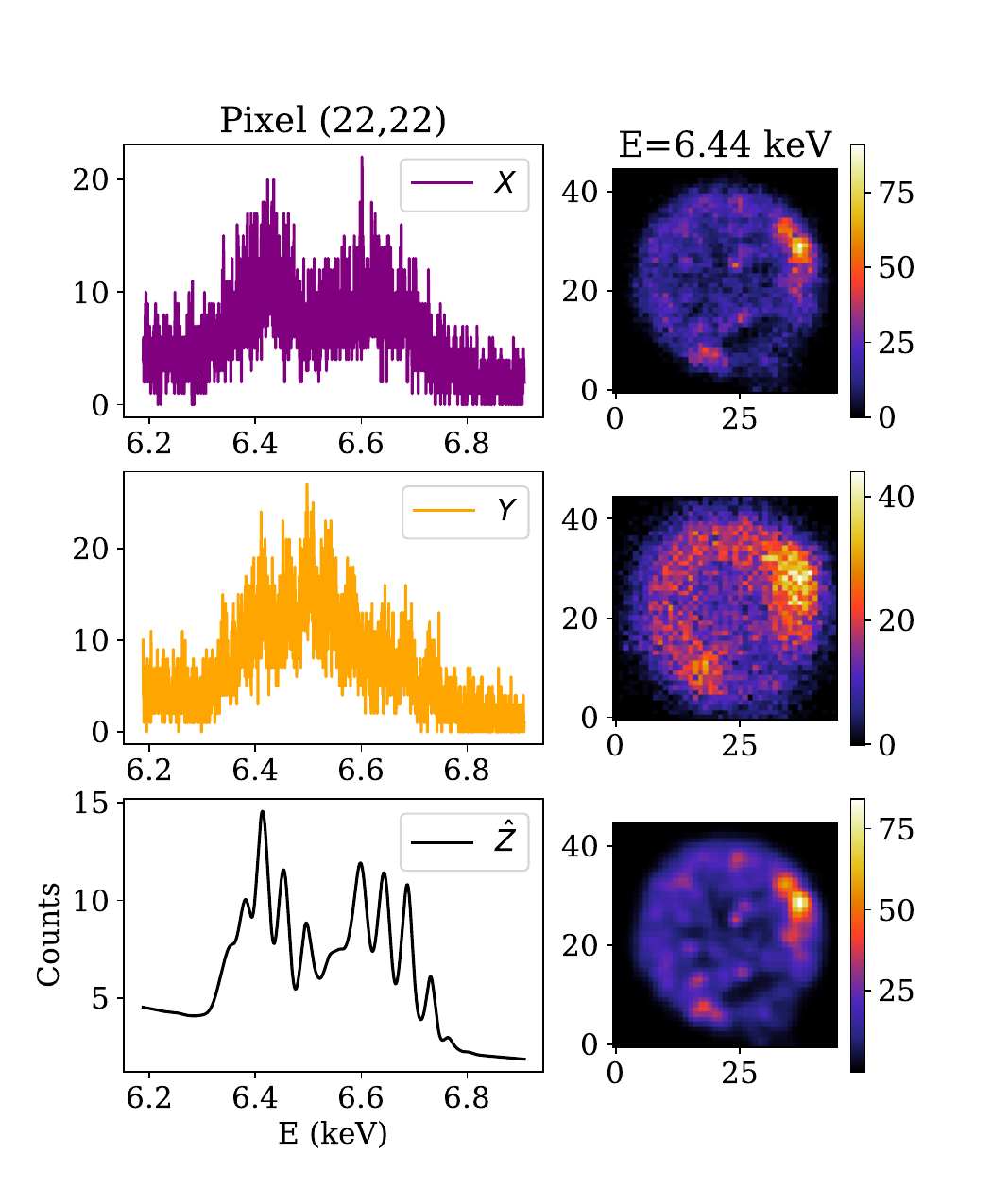}
    \caption{}
\end{subfigure}
\\
\begin{subfigure}{0.48\textwidth}
      \centering
    \includegraphics[width=\textwidth]{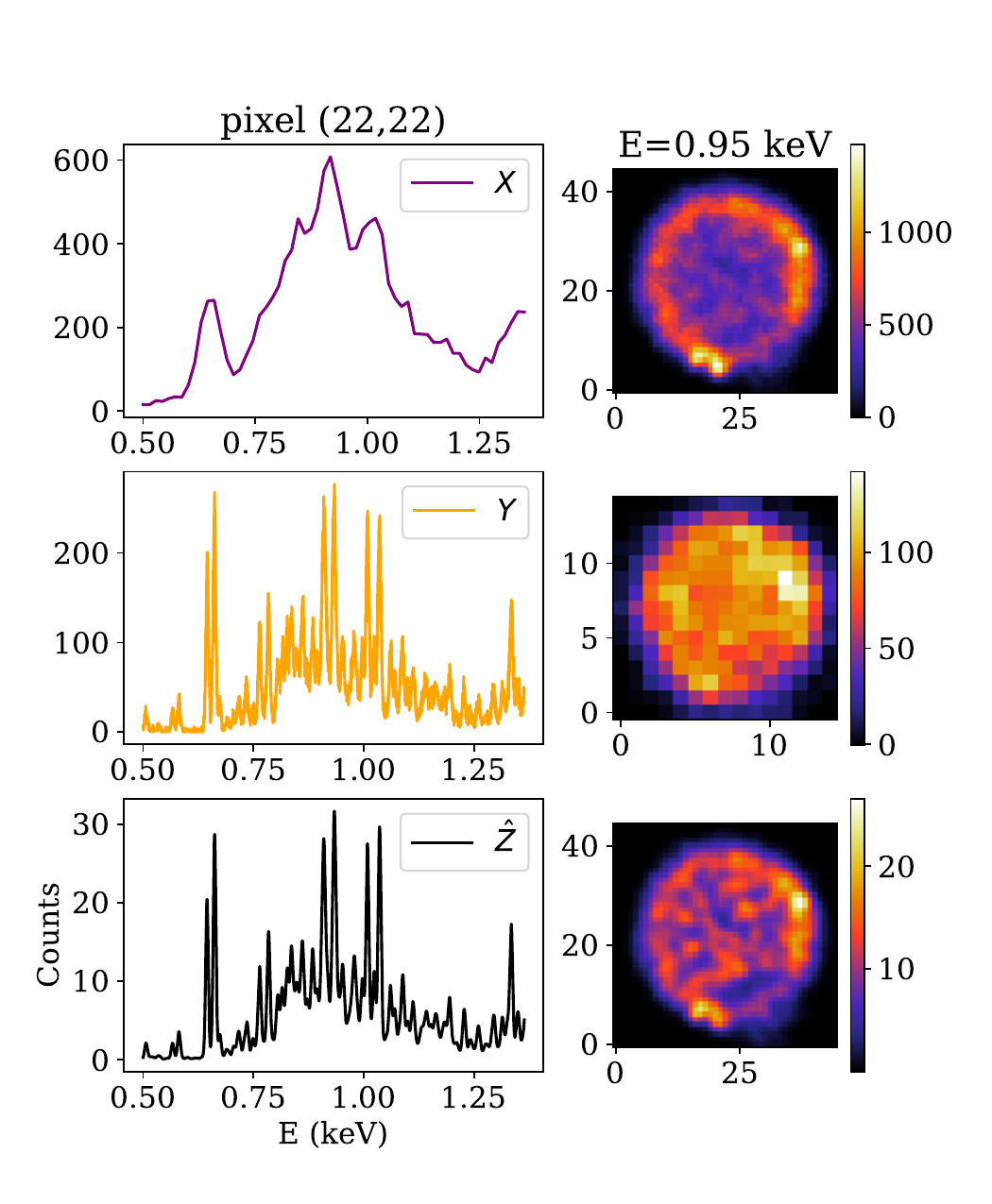}
    \caption{}
\end{subfigure}
\begin{subfigure}{0.48\textwidth}
      \centering
    \includegraphics[width=\textwidth]{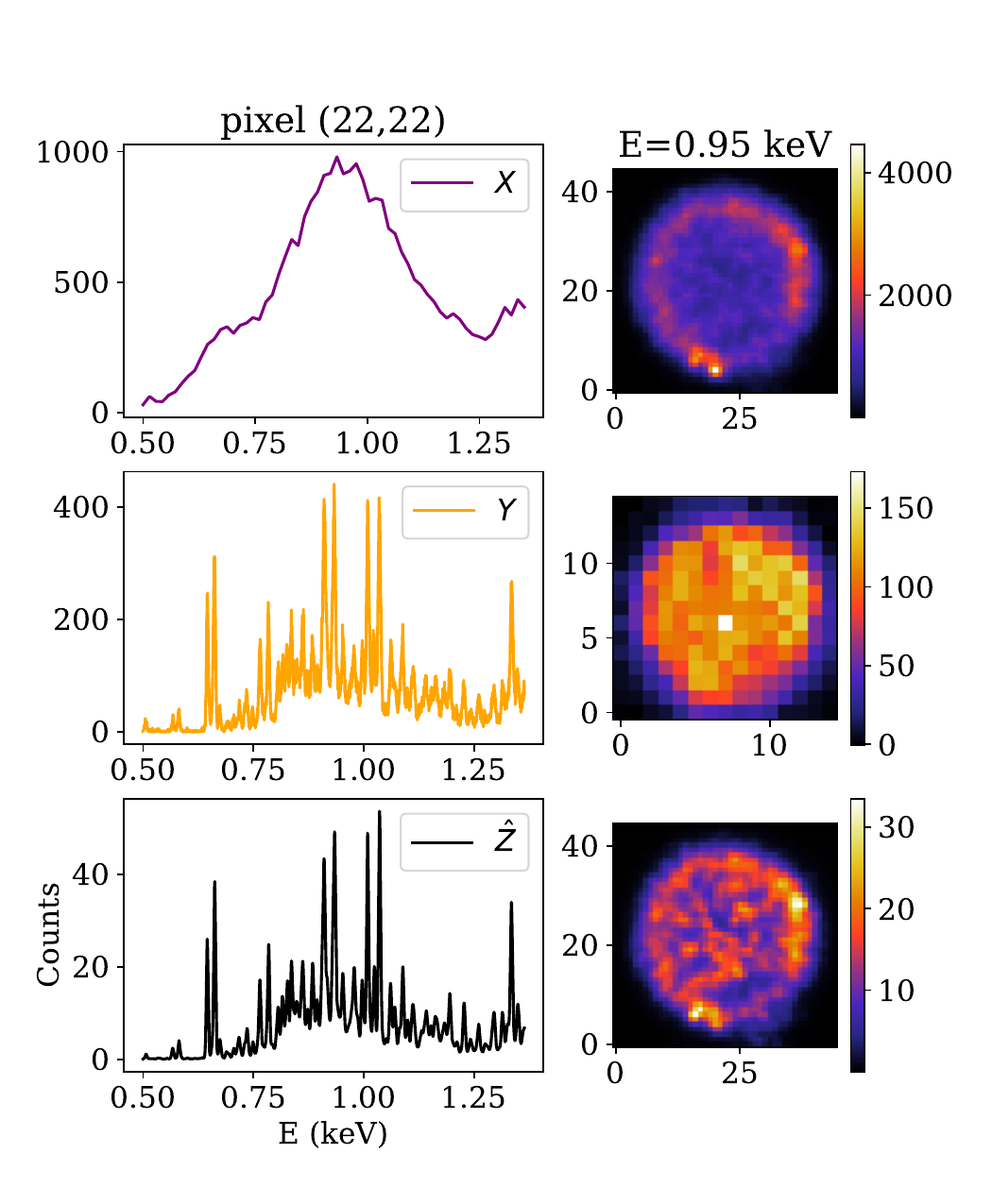}
    \caption{}
\end{subfigure}
\caption{Example of spectra and 2D slices from the four toy-model data sets.  The ground truth $\hat{Z}$ is shown in black, and the two data sets in purple and orange: $X$ with the higher spatial resolution, and $Y$ with the higher spectral resolution. \\
(a) Gaussian model between 0.5 and 1.4 keV, (b) Gaussian model between 6.2 and 6.9 keV, (c) Gaussian model with rebinning between 0.5 and 1.4 keV, and (d) the Realistic model between 0.5 and 1.4 keV. } 

\label{fig:example_toymodels}
\end{figure*}

In all of these case studies, we generated toy models with a photon count of around $10^7$. We then applied Poisson noise. Some sample spectra and 2D slices are shown in Fig. \ref{fig:example_toymodels}. 

\subsection{Results}
\label{sec:results_plots}

This section presents the results we obtained with the four toy models. We compare the wavelet 2D-1D regularization (W2D1D), the low-rank approximation with Sobolev regularization (LRS), and the low-rank approximation with wavelet 2D regularization (LRW2D). We present our results in the form of sample pixel fits, amplitude maps, error histograms,  error maps, and scalar metrics. 

\begin{figure*}[h!]
\centering
\begin{subfigure}{0.4\textwidth}
    \includegraphics[width=\textwidth]{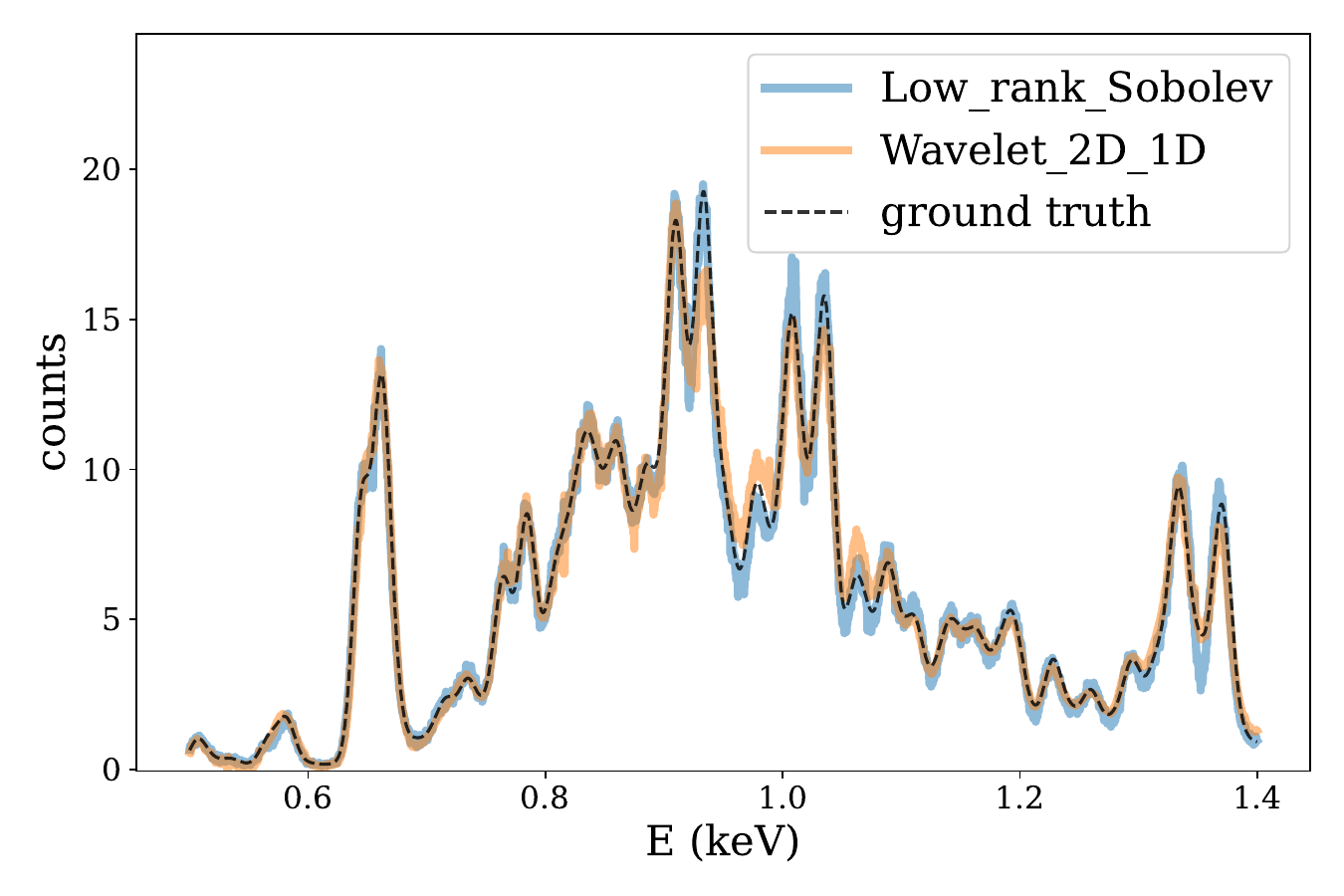}
    \caption{}
\end{subfigure}
\begin{subfigure}{0.4\textwidth}
    \includegraphics[width=\textwidth]{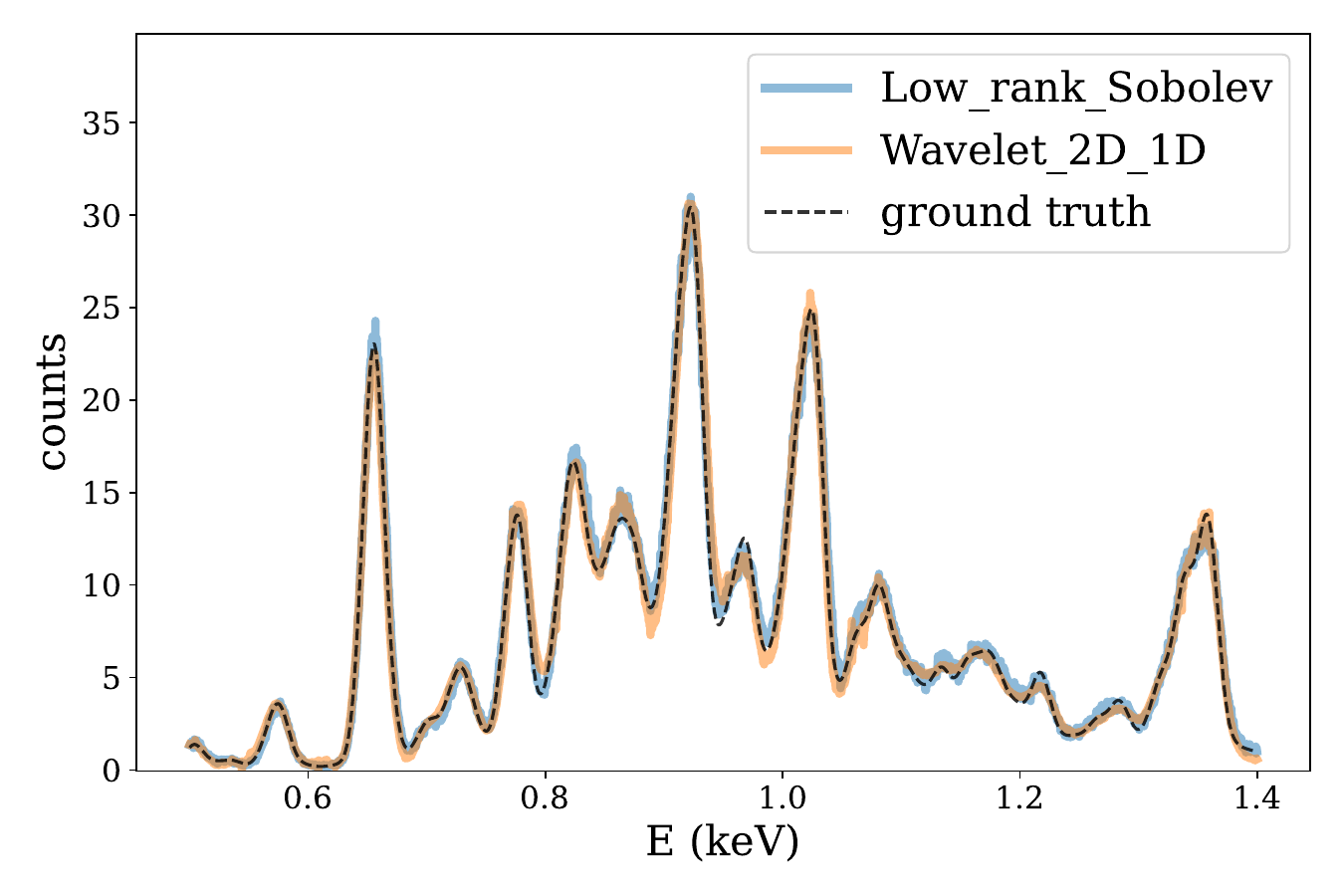}
    \caption{}
\end{subfigure}
\\
\begin{subfigure}{0.4\textwidth}
    \includegraphics[width=\textwidth]{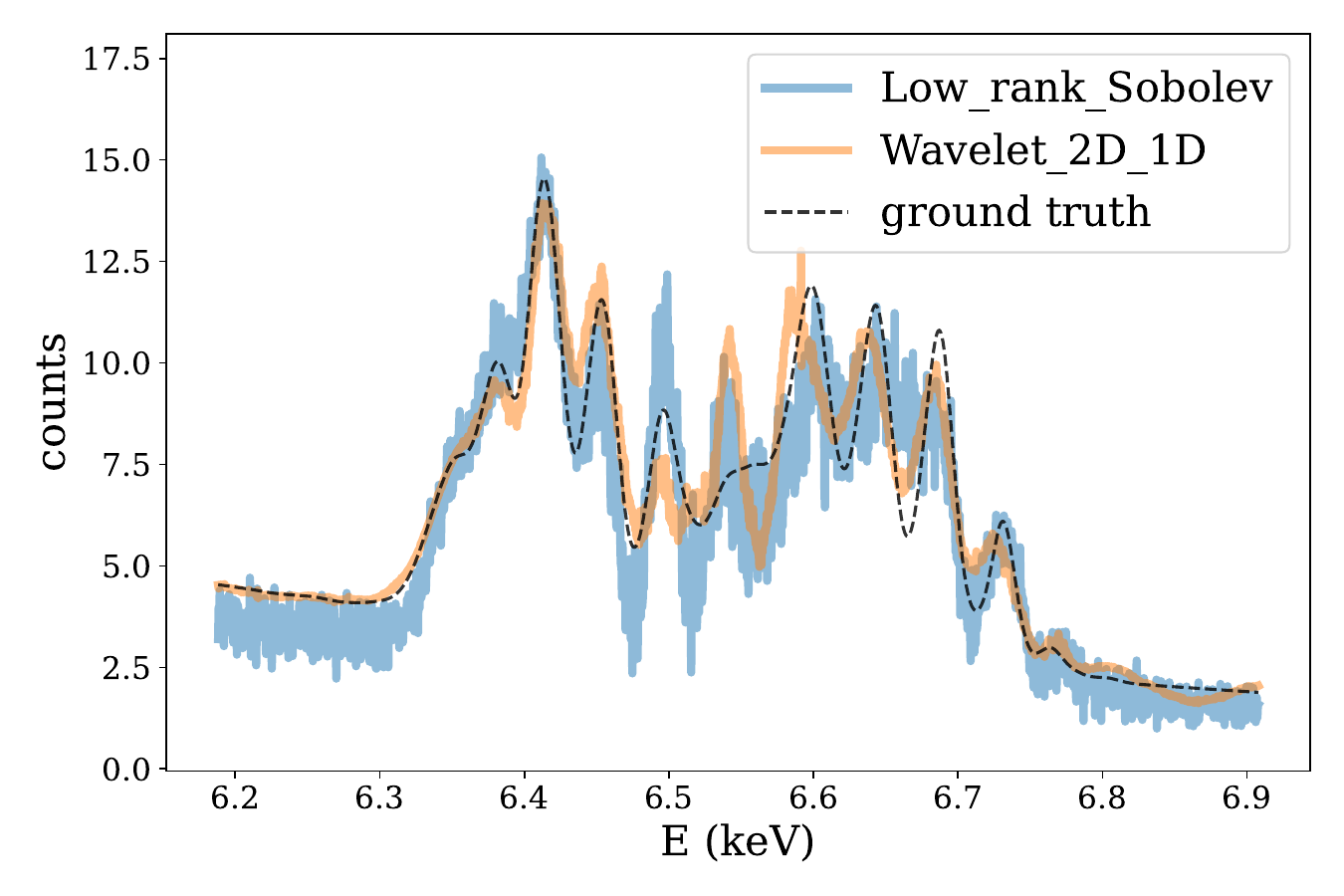}
    \caption{}
\end{subfigure}
\begin{subfigure}{0.4\textwidth}
    \includegraphics[width=\textwidth]{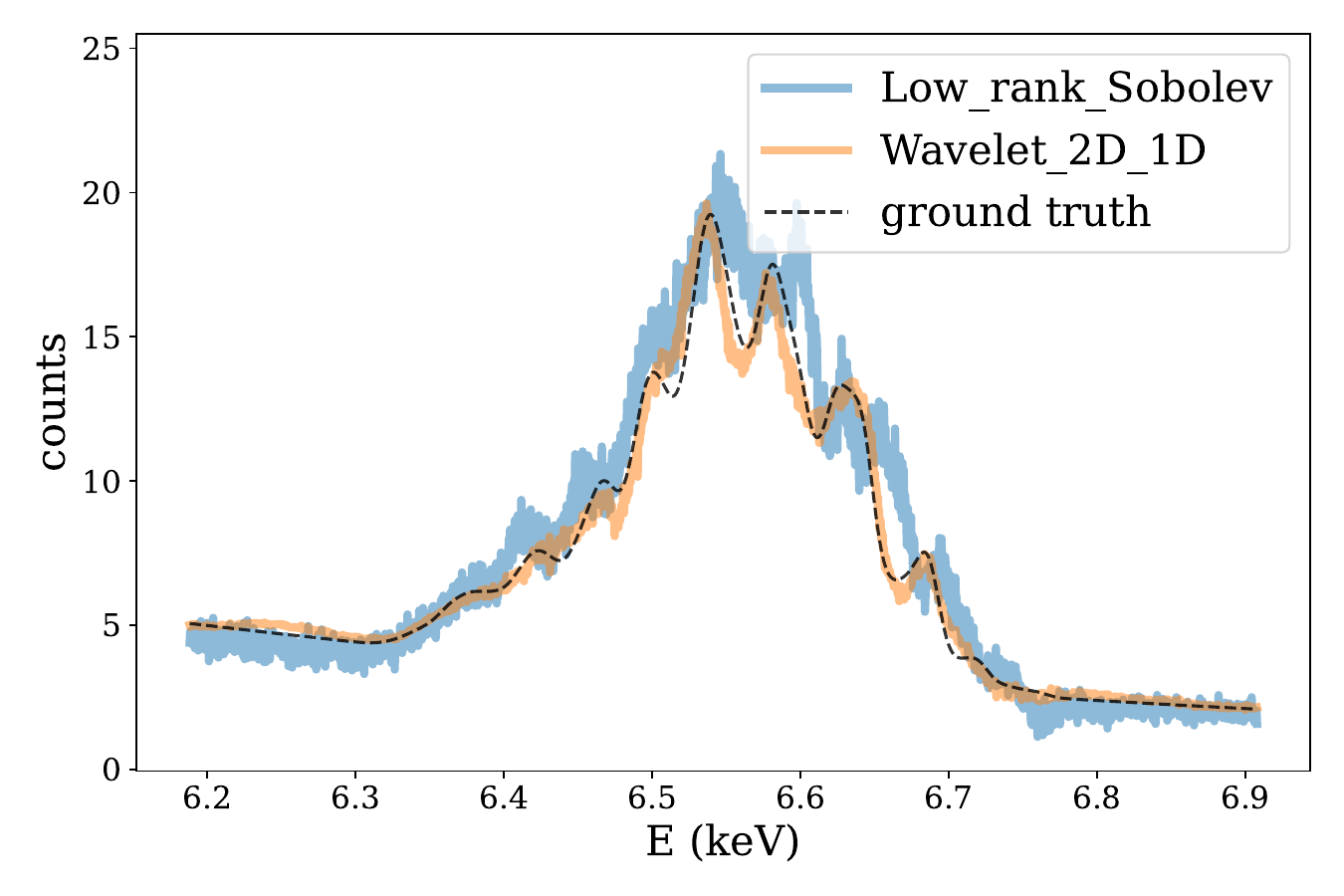}
    \caption{}
\end{subfigure}
\\
\begin{subfigure}{0.4\textwidth}
    \includegraphics[width=\textwidth]{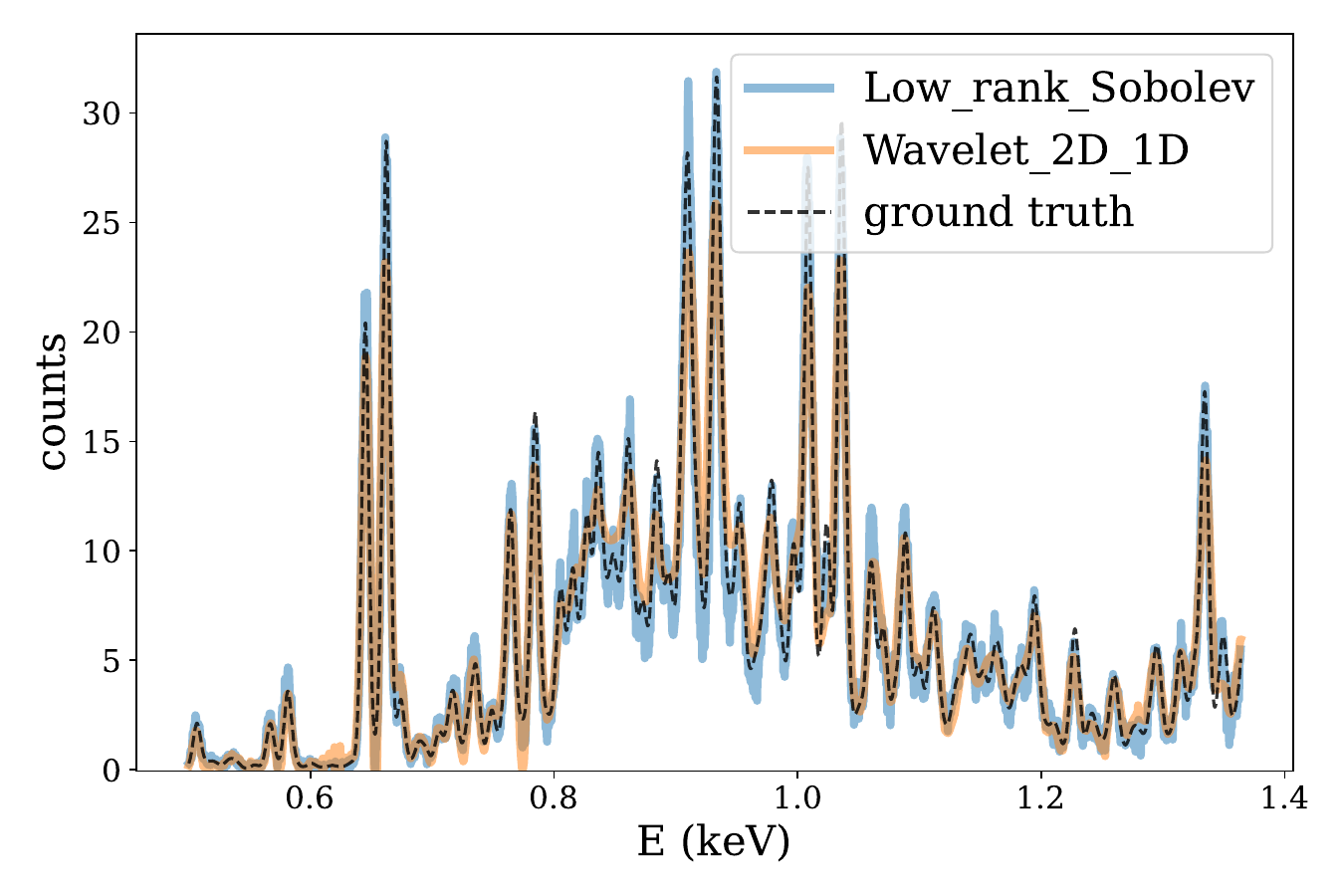}
\end{subfigure}
\begin{subfigure}{0.4\textwidth}
    \includegraphics[width=\textwidth]{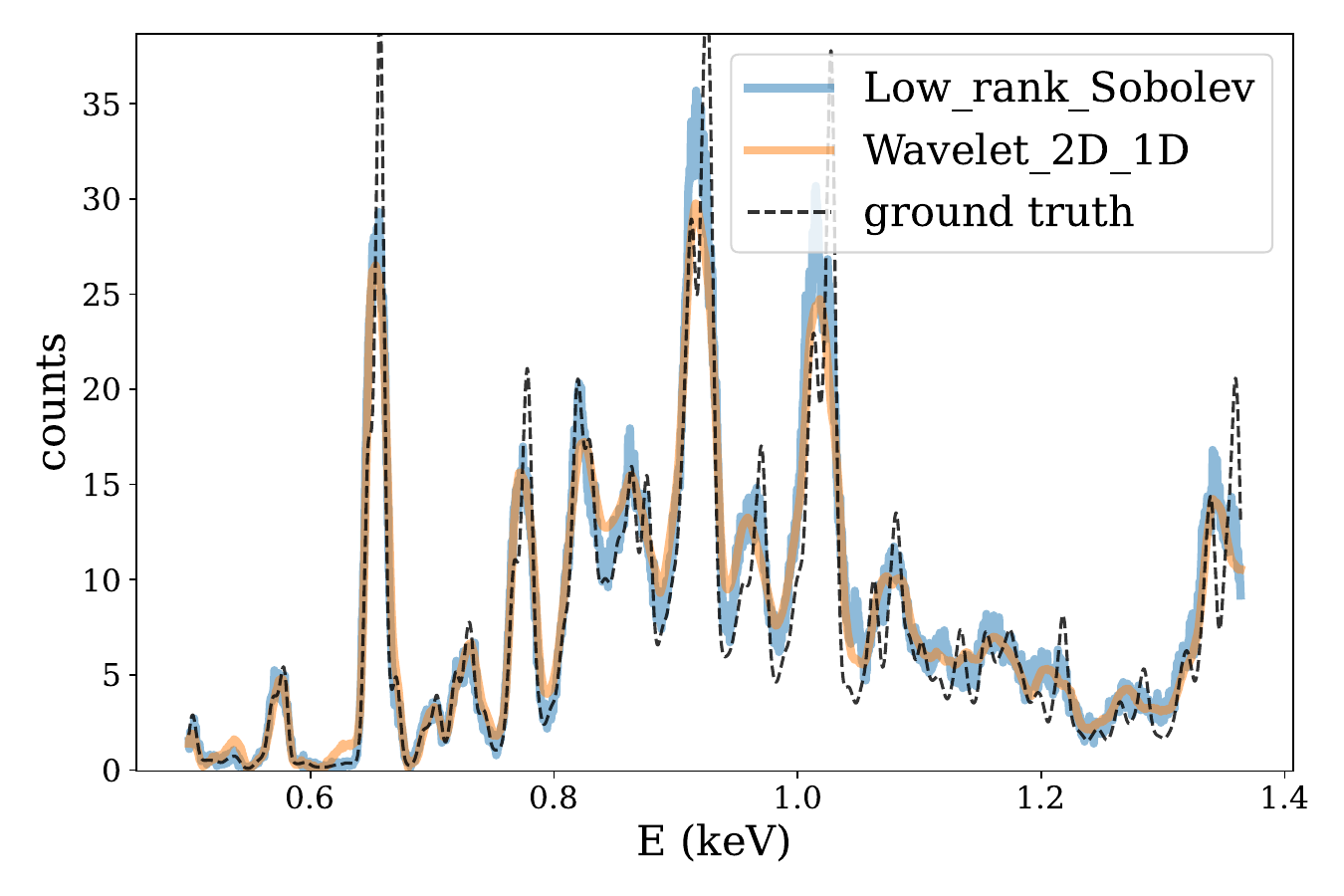}
\end{subfigure}
\\
\begin{subfigure}{0.4\textwidth}
    \includegraphics[width=\textwidth]{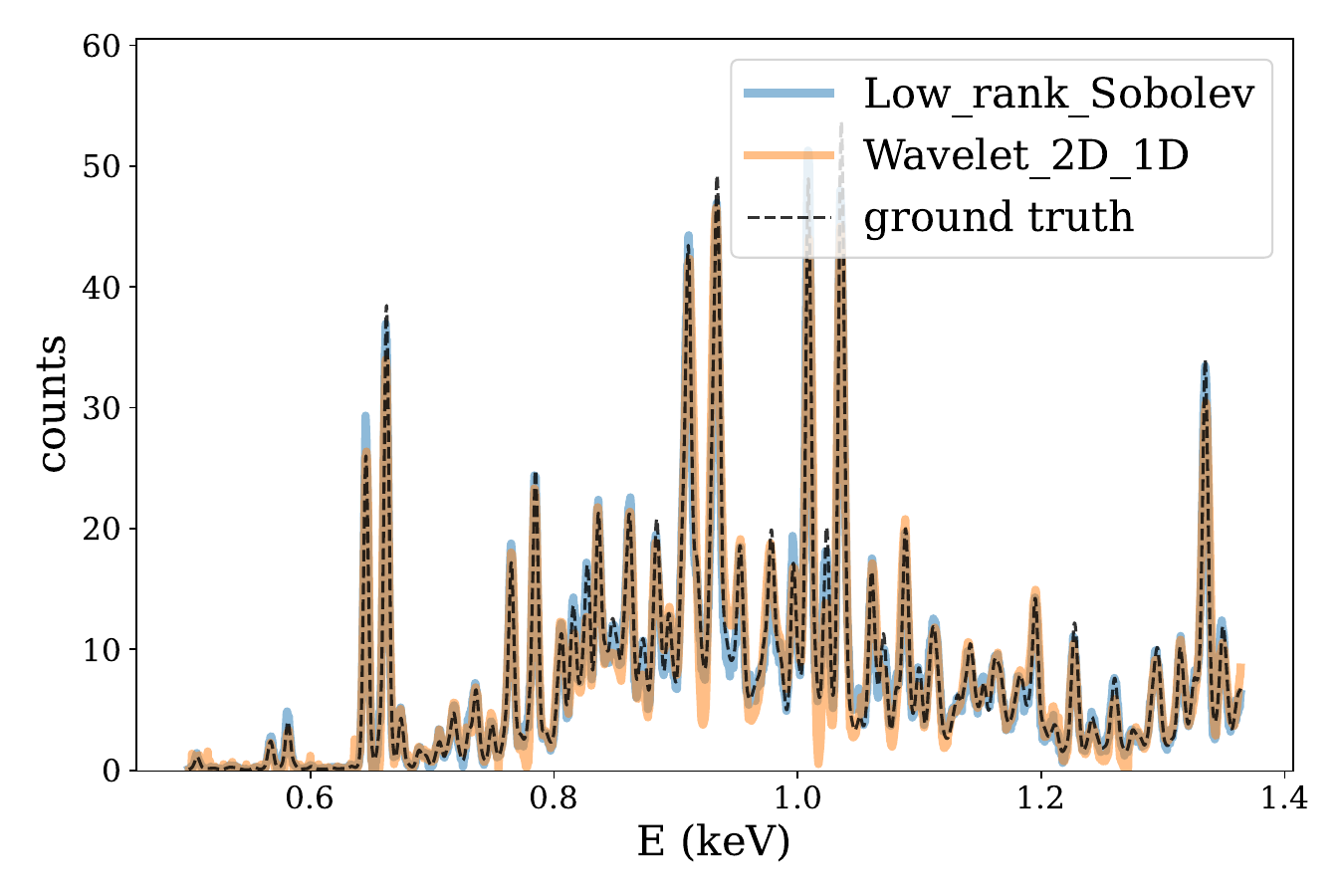}
    \caption{}
\end{subfigure}
\begin{subfigure}{0.4\textwidth}
    \includegraphics[width=\textwidth]{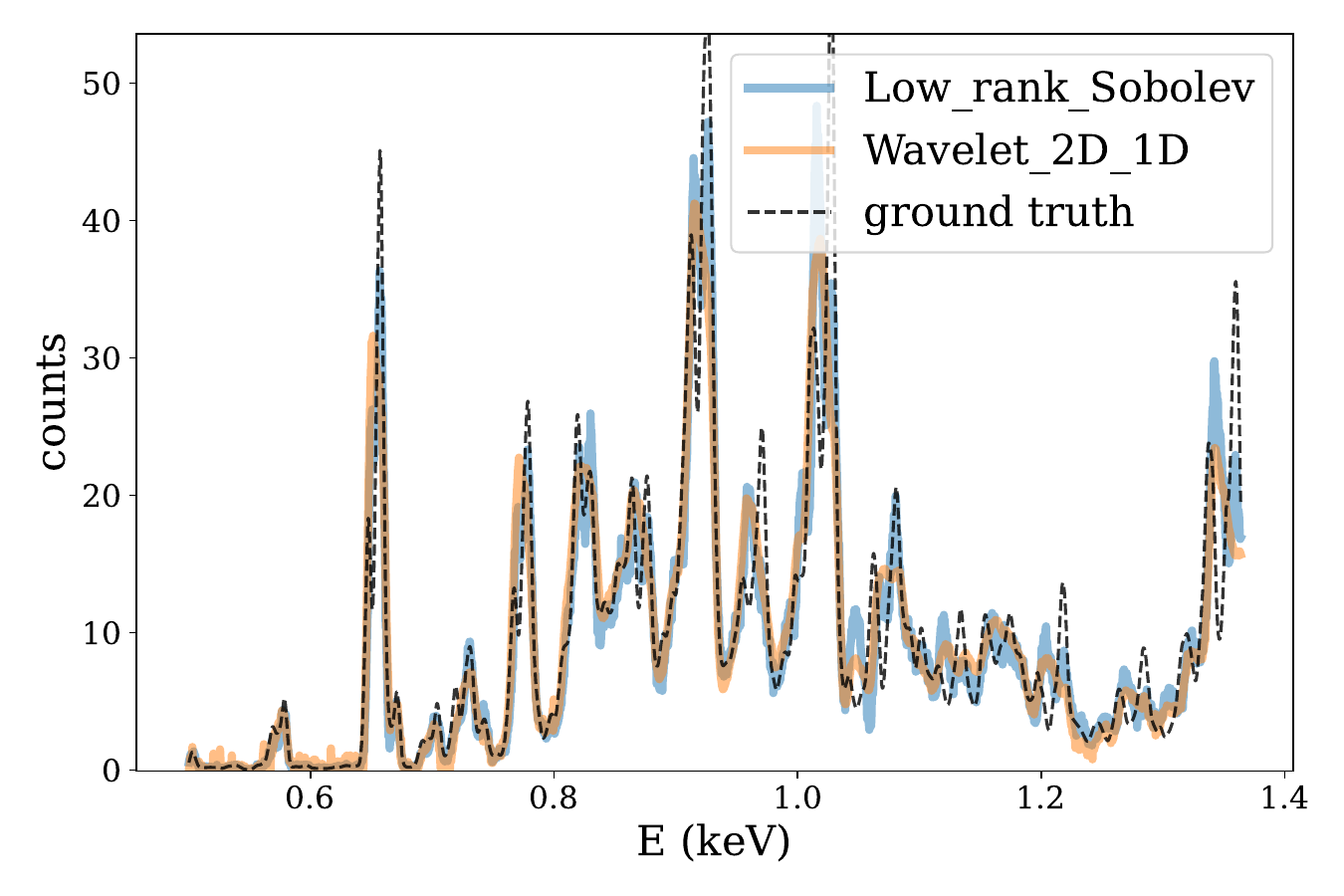}
    \caption{}
\end{subfigure}
\vspace{5 pt}
\caption{Example of how the fusion output compares to the ground truth for two pixels. From top to bottom, the models are: (a,b) Gaussian model between 0.5 and 1.4 keV, (c,d) Gaussian model between 6.2 and 6.9 keV, (e,f) Gaussian model with rebinning, and (g,h) the Realistic model. 
The left column shows results for pixel (22,22), which is lower in brightness than pixel (11,31) on the right. }
\label{fig:example_pixel_fits}
\end{figure*}

First, the reconstructed spectra obtained on two sample pixels are shown in Figure \ref{fig:example_pixel_fits}. Once again, for visual clarity, we only show the W2D1D and LRS methods. These results allow us to draw several conclusions. First of all, the Gaussian model around the iron line (c, d) is difficult to reconstruct for both methods, but the LRS method appears more biased. Second, we see that the W2D1D method results in smoother spectra, but this can be accompanied by more bias, as seen in the Gaussian model with rebinning (e,f). This occurs when the underlying assumption of sparsity in the wavelet domain is not appropriate. 

The amplitude maps are displayed in Fig. \ref{fig:amp_maps}. Overall, for all regularizations, we see that the method is able to reconstruct the general spatial features of the remnant. 
For the Gaussian model with rebinning, we do notice some amount of leakage on the edges of the remnant for all regularizations, particularly on the western side. For the Gaussian model around 6 keV (b), we also see that LRS has over-smoothed some of the features — this effect is visible in the other toy models but is more subtle. Finally, in models with rebinning (c, d), the LRW2D seems to smooth out finer features and give a biased result.

In Fig. \ref{fig:spectral_angle_err}, we present the map of the spectral angle mapper (SAM), an error metric defined as (for a spectra $Z_i$)
\begin{equation}
    SAM(Z_i)=\arccos \Bigg( \frac{\langle Z_i, \hat{Z}_i \rangle}{||Z_i||_2||\hat{Z}_i||_2} \Bigg).
    \label{eq:sam}
\end{equation}
The SAM is best if low (close to zero), and is designed to capture spectral distortions. In Fig. \ref{fig:spectral_angle_err}, we observe some leakage on the low-count pixels at the edges of the remnant, which is present for all methods, but more so for W2D1D. The results shown in panel (a) suggest that the wavelet regularization presents slightly more spatial distortion for the simplest model (Gaussian model with low spectral variability), while it appears more successful in panel (b) for the Gaussian model with high spectral variability. In panel (b), the higher SAM pixels in LRS and LRW2D correspond to some local features, in areas of high spectral variability (see the middle of Fig. \ref{fig:Angular_distance_TM}); we highlight, for instance, the cluster of bright spots above the center of the remnant. In panel (c), the low rank with Sobolev appears to be doing better in the center. All models display a high error close to the bright spot south of the remnant. 
In panel (d), for the realistic model, LRS obtains better result in the center of the remnant. While all three methods struggle to reach a result close to the ground truth in the bright area to the south of a remnant, LRW2D and W2D1D  show the poorest performance, which could suggest that the Sobolev regularization is better at handling outliers. As this bright spot is also seen as a problem in the model shown in panel (c), this area of high flux and high spatial variability seems especially difficult to reconstruct in models with rebinning. 

In Fig. \ref{fig:err_percent}, we display the relative error maps averaged per pixel. We draw similar conclusions about the Gaussian model without rebinning. The difficulty that low-rank methods have in reconstructing high spectral variabilities areas is once again seen in panel (b). For panel (c), we again notice the leakage on the edges, with the amplitude being judged too high in red areas, and too low in blue areas. There is a general bias on all models that seems to overestimate the northwest part of the remnant, while underestimating the southeast part. This is likely due to the rebinning, as it is also present in panel (d).
Further, in panel (d), we note that W2D1D and LRW2D underestimated the amplitude of the bright spot in the south of the remnant, while LRS was not as biased. Overall, W2D1D displays more bias in the southern bright outlier than the other methods.

Finally, we measured scalar error metrics. All results can be found in Table \ref{tab:error_metrics}. The best results are highlighted in bold. As \cite{zhang2008} showed for the case of fusion between panchromatic and MS images, scalar metrics do not necessarily agree when comparing the efficacy of fusion methods, and are not always easily interpretable. Thus, we present four metrics in order to provide a more complete comparison. The definitions of the metrics can be found in Appendix \ref{sec:metrics}. Namely, we calculate the normalised mean squared error (NMSE), the average spectral angle mapper (aSAM), average complementary structural similarity (acSSIM) index, and the relative dimensionless global error (ERGAS).
\begin{table}
    \centering
    \caption{Error metric results comparing regularizations for the four toy models.}
    \begin{tabular}{|c|c|c|c|c|}\hline
 \multicolumn{5}{|c|}{\textbf{Gaussian (0.5-1.4 keV)}}\\\hline 
         \emph{Regularization}&  NMSE&  aSAM& acSSIM & ERGAS\\ \hline 
         W2D1D&  \textbf{21.36}&  0.571& 0.017 & 0.220\\ \hline 
 LRS& 20.97& \textbf{0.557}&\textbf{0.011}& 0.116\\ \hline 
 LRW2D& 21.01& 0.560&0.012 & \textbf{0.114} \\\hline 
         \multicolumn{5}{|c|}{\textbf{Gaussian (6.2-6.9 keV)}}\\ \hline 
 & NMSE& aSAM&acSSIM& ERGAS\\ \hline 
 W2D1D& \textbf{18.67}& \textbf{0.599}&\textbf{0.005}& \textbf{0.229} \\ \hline 
 LRS&   14.16& 0.600&0.011& 0.343 \\ \hline 
 LRW2D&         16.57& 0.609&0.010& 0.248\\ \hline 
         \multicolumn{5}{|c|}{\textbf{Gaussian with Rebinning (0.5-1.4 keV)}}\\ \hline 
 & NMSE& aSAM&acSSIM& ERGAS\\ \hline 
 W2D1D& 10.36& 0.691 & 0.048 & 0.422 \\ \hline 
 LRS& \textbf{11.01} & \textbf{0.670} & \textbf{0.045} & 0.341 \\ \hline 
 LRW2D& 10.34&          0.690 & 0.054 & \textbf{0.317}\\ \hline 
         \multicolumn{5}{|c|}{\textbf{Realistic with Rebinning (0.5-1.4 keV)}}\\ \hline 
 & NMSE& aSAM&acSSIM& ERGAS\\ \hline 
 W2D1D& 9.956 & 0.700 & 0.032 & 0.608 \\ \hline 
 LRS& \textbf{10.16} & \textbf{0.691} & \textbf{0.031} & 0.358 \\ \hline
 LRW2D& 9.523 & 0.712 & 0.038 & \textbf{0.351}\\ \hline
    \end{tabular}
    \tablefoot{The NMSE (Equation \ref{eq:NMSE}) is better if high, while the aSAM (Equation \ref{eq:aSAM}) and acSSIM (Equation \ref{eq:acSSIM}) are better if low. The best values are marked in bold.}
    \label{tab:error_metrics}
\end{table}

Looking at the obtained values in Table \ref{tab:error_metrics}, we see that the four metrics rarely draw the same conclusion, except in the case of the Gaussian model around 6 keV, which is the model with the highest spectral variability between pixels. This allows us to draw a similar conclusion as for the error maps: the methods using a low-rank approximation have more trouble accounting for high spectral variability. This may be due to the fact that the low-rank approximation is learnt by performing PCA on $Y$, the data set that has the best spectral resolution, but its poor spatial resolution will smooth  out spectral variability between pixels. If the low rank could be learnt on $Z$, this would not be a problem, but of course that is not possible. 

The results on the Gaussian model around 1 keV show that all three methods are the best in at least one metric, and so it is hard to choose an outright leader. In this case, low rankedness does seem to lower spectral distortion (looking at the aSAM and the ERGAS). For the Gaussian model around 1 keV with rebinning, the low-rank methods are more successful, but the difference is small, and the bright southern spot we note on the maps likely plays a part in the W2D1D being a little less successful. We also observe this for the realistic model with rebinning. 
Overall, it appears that the low-rank methods are best at dealing with outliers, especially the low rank with Sobolev, but they are not as appropriate for models with high spectral variability. 

Convergence took between 30k and 100k iterations, depending on the toy model. On average, one iteration of the algorithm took around 3s with the method with 2D-1D wavelet regularization, resulting in a computation time of between 25h and 83h. With the low-rank approximation with Sobolev, each iteration took around 2s, resulting in a computation time of between 15h and 55h; the low rank with wavelet 2D was around the same. 

\section{Discussion}
\label{sec:discussion}
\paragraph{}
Overall, the wavelet 2D-1D regularization is more successful than the low-rank approximations at dealing with  strong spectral variability across the image. However, the assumption of wavelet sparsity is not necessarily ideal to account for certain structures, such as in the presence of singular bright spots like the one in the southern part of the remnant, and this is especially difficult to treat with the inclusion of a rebinning operator. Nonetheless, we find that for all three regularization methods, the fusion results are satisfactory given the difficulty of the problem; however, they still contain some bias. 

The low-rank approximation, while ill-suited for SNRs that display strong spectral variability due to the variations of their physical parameters, was found to accelerate the algorithm by half thanks to the dimension reduction. As the data sets are large and the algorithm takes a long time to converge, this is a useful attribute. 

Thus, it would be ideal to find a spectral regularization less biased than the wavelet sparsity or the low-rank approximation, but that nonetheless includes a reduction in dimensions. This may be done by exploring a low-rank approximation more refined than one learnt using PCA on the high-spectral-resolution data set. We could also explore acceleration methods, such as algorithm unrolling, as in \cite{fahes2022unrolling}. 

Alternatively, a promising avenue would be to include physical information. Indeed, while having a blind fusion result is useful (as it requires no assumptions on the model), in order to obtain physical parameter maps, a model will need to be fitted to the result. In the case of SNRs, more than one model would be needed, as the emission comes from a mixture composed of thermal radiation from the shocked interstellar media and synchrotron radiation due to particle acceleration. 

An algorithm for source separation that includes spectral variations was developed by \cite{Sushi_article}. Our future work will focus on combining this approach to \texttt{HIFReD}, we the aim being to develop a method capable of simultaneously fusing data sets from two generations of telescopes and perform unmixing with a physical spectral model.

On a final note, one limitation of the proposed method is that as it is implemented, it cannot account for spatially varying PSF, or a spectrally varying spectral response. In this case, the convolution is no longer a term-by-term product in Fourier. This is an issue with XMM/Resolve, where a constant spectral response over a wide energy range is not realistic. Therefore, we are currently working on developing a version of HIFReD than can handle varying convolution kernels, with promising results so far. The main obstacle is the computational complexity, which is greatly increased, but by working on an accelerated version of the code using the JAX\footnote{\url{https://jax.readthedocs.io/}} Python package, we find that we should be able to achieve even faster performances than those described in this article, even when using a non-stationary convolution. 
\section{Conclusions}
\paragraph{}
We have developed an algorithm for fusing hyperspectral images taken by two generations of X-ray telescopes with complementary spatial and spectral resolutions. Our method simultaneously deconvolves the instruments' spatial and spectral responses in order to return to the best resolutions of the two telescopes.

We implemented three regularization schemes, the first being a term to promote sparsity under a 2D-1D wavelet transform, the second being a low-rank approximation with Sobolev regularization  \citep[inspired by ][]{guilloteau_2020}, and the third being a low-rank approximation with a 2D wavelet sparsity term. 

We tested our method on five toy models constructed from hydrodynamic simulations of SNRs by \cite{Orlando_2016} and convolved with an increasing complexity of instrumental response. Spatial features are well reconstructed, apart from a slight leakage in the bright shell in the toy model with rebinning. The reconstruction of spectral features depends on the regularizations and model.

We find that while low-rank methods accelerated the algorithm by half, they struggled to account for the strong spectral variability present around the iron line. The 2D-1D wavelet method  better accounts for this variability, but it is somewhat biased in reconstructing the realistic model. 

In conclusion, we believe that hyperspectral fusion shows promise and would greatly benefit the analysis of XRISM/Resolve data; though it would be useful to explore other regularization terms more suited to the variabilities of the data. In particular, we intend to combine fusion with a spectral physical model and source-separation algorithm in future works. While the purpose of this article is to present a proof of concept, we are now actively working on applying this method to real data. 

\begin{acknowledgements}
The research leading to these results has received funding from the European Union’s Horizon 2020 Programme under the AHEAD2020 project (grant agreement n. 871158). This work was supported by CNES, focused on methodology for X-ray analysis. 
This work was supported by the Action Thématique Hautes Energies of CNRS/INSU with INP and IN2P3, co-funded by CEA and CNES. We thank S. Orlando for kindly providing the simulation from \cite{Orlando_2016}  to generate our toy model.
\end{acknowledgements}

\bibliographystyle{aa}
\bibliography{bibliography}

\begin{thebibliography}{25}
\expandafter\ifx\csname natexlab\endcsname\relax\def\natexlab#1{#1}\fi

\bibitem[{Bach {et~al.}(2011)Bach, Jenatton, Mairal, \& Obozinski}]{sparsity}
Bach, F., Jenatton, R., Mairal, J., \& Obozinski, G. 2011, Foundations and Trends in Machine Learning, 4, 1

\bibitem[{{Barret} {et~al.}(2018){Barret}, {Lam Trong}, {den Herder}, {Piro}, {Cappi}, {Houvelin}, {Kelley}, {Mas-Hesse}, {Mitsuda}, {Paltani}, {Rauw}, {Rozanska}, {Wilms}, {Bandler}, {Barbera}, {Barcons}, {Bozzo}, {Ceballos}, {Charles}, {Costantini}, {Decourchelle}, {den Hartog}, {Duband}, {Duval}, {Fiore}, {Gatti}, {Goldwurm}, {Jackson}, {Jonker}, {Kilbourne}, {Macculi}, {Mendez}, {Molendi}, {Orleanski}, {Pajot}, {Pointecouteau}, {Porter}, {Pratt}, {Pr{\^e}le}, {Ravera}, {Sato}, {Schaye}, {Shinozaki}, {Thibert}, {Valenziano}, {Valette}, {Vink}, {Webb}, {Wise}, {Yamasaki}, {Douchin}, {Mesnager}, {Pontet}, {Pradines}, {Branduardi-Raymont}, {Bulbul}, {Dadina}, {Ettori}, {Finoguenov}, {Fukazawa}, {Janiuk}, {Kaastra}, {Mazzotta}, {Miller}, {Miniutti}, {Naze}, {Nicastro}, {Scioritino}, {Simonescu}, {Torrejon}, {Frezouls}, {Geoffray}, {Peille}, {Aicardi}, {Andr{\'e}}, {Daniel}, {Cl{\'e}net}, {Etcheverry}, {Gloaguen}, {Hervet}, {Jolly}, {Ledot}, {Paillet}, {Schmisser}, {Vella}, {Damery}, {Boyce}, {Dipirro},
  {Lotti}, {Schwander}, {Smith}, {Van Leeuwen}, {van Weers}, {Clerc}, {Cobo}, {Dauser}, {Kirsch}, {Cucchetti}, {Eckart}, {Ferrando}, \& {Natalucci}}]{Barret2018}
{Barret}, D., {Lam Trong}, T., {den Herder}, J.-W., {et~al.} 2018, in Society of Photo-Optical Instrumentation Engineers (SPIE) Conference Series, Vol. 10699, Space Telescopes and Instrumentation 2018: Ultraviolet to Gamma Ray, ed. J.-W.~A. {den Herder}, S.~{Nikzad}, \& K.~{Nakazawa}, 106991G

\bibitem[{Bertero~M. \& C.(2022)}]{Bertero_2022}
Bertero~M., B.~P. \& C., D.~M. 2022, Introduction to Inverse Problems in Imaging (CRC Press)

\bibitem[{Fahes {et~al.}(2022)Fahes, Kervazo, Bobin, \& Tupin}]{fahes2022unrolling}
Fahes, M., Kervazo, C., Bobin, J., \& Tupin, F. 2022, in International Conference on Learning Representations

\bibitem[{Godinaud {et~al.}(2023)Godinaud, Acero, Decourchelle, \& Ballet}]{Godinaud_2023}
Godinaud, L., Acero, F., Decourchelle, A., \& Ballet, J. 2023, Astronomy and Astrophysics, 680, A80

\bibitem[{Guilloteau {et~al.}(2022)Guilloteau, Oberlin, Bern{\'e}, \& Dobigeon}]{guilloteau_spatreg_2022}
Guilloteau, C., Oberlin, T., Bern{\'e}, O., \& Dobigeon, N. 2022, in {IEEE International Conference on Image Processing (ICIP 2022)}, Bordeaux, France, 1--5

\bibitem[{Guilloteau {et~al.}(2020)Guilloteau, Oberlin, Berné, Émilie Habart, \& Dobigeon}]{guilloteau_2020}
Guilloteau, C., Oberlin, T., Berné, O., Émilie Habart, \& Dobigeon, N. 2020, The Astronomical Journal, 160, 28

\bibitem[{Hadamard(1923)}]{Hadamard}
Hadamard, J. 1923, Lectures on the Cauchy Problem in Linear Partial Differential Equations (New Haven: Yale University Press)

\bibitem[{Karl(2005)}]{Sobolev}
Karl, W.~C. 2005, in LHandbook of Image and Video Processing. Elsevier (Elsevier)

\bibitem[{{Lascar, J.} {et~al.}(2024){Lascar, J.}, {Bobin, J.}, \& {Acero, F.}}]{Sushi_article}
{Lascar, J.}, {Bobin, J.}, \& {Acero, F.} 2024, Astronomy and Astrophysics, 686, A259

\bibitem[{Orlando {et~al.}(2016)Orlando, Miceli, Pumo, \& Bocchino}]{Orlando_2016}
Orlando, S., Miceli, M., Pumo, M.~L., \& Bocchino, F. 2016, The Astrophysical Journal, 822, 22

\bibitem[{Parikh \& Boyd(2014)}]{prox}
Parikh, N. \& Boyd, S. 2014, Foundations and Trends in Optimization, 1, 127–239

\bibitem[{Pineau {et~al.}(2023)Pineau, Orieux, \& Abergel}]{pineau_2023}
Pineau, D., Orieux, F., \& Abergel, A. 2023, working paper or preprint

\bibitem[{Pr{\'e}vost {et~al.}(2022)Pr{\'e}vost, Borsoi, Usevich, Brie, Bermudez, \& Richard}]{prevost2022}
Pr{\'e}vost, C., Borsoi, R.~A., Usevich, K., {et~al.} 2022, {SIAM Journal on Imaging Sciences}, 15, 110

\bibitem[{Simões {et~al.}(2015)Simões, Bioucas-Dias, Almeida, \& Chanussot}]{HySure}
Simões, M., Bioucas-Dias, J., Almeida, L.~B., \& Chanussot, J. 2015, IEEE Transactions on Geoscience and Remote Sensing, 53, 3373

\bibitem[{{Starck} \& {Murtagh}(1994)}]{starlet}
{Starck}, J.-L. \& {Murtagh}, F. 1994, Astronomy and Astrophysics, 342

\bibitem[{Starck {et~al.}(2010)Starck, Murtagh, \& Fadili}]{starckbook}
Starck, J.-L., Murtagh, F., \& Fadili, Jalal, M. 2010, Sparse Image and Signal Processing: Wavelets, Curvelets, Morphological Diversity (Cambridge University Press)

\bibitem[{{Takahashi} {et~al.}(2014){Takahashi}, {Mitsuda}, {Kelley}, {Aharonian}, {Akamatsu}, {Akimoto}, {Allen}, {Anabuki}, {Angelini}, {Arnaud}, {Asai}, {Audard}, {Awaki}, {Azzarello}, {Baluta}, {Bamba}, {Bando}, {Bautz}, {Bialas}, {Blandford}, {Boyce}, {Brenneman}, {Brown}, {Cackett}, {Canavan}, {Chernyakova}, {Chiao}, {Coppi}, {Costantini}, {de Plaa}, {den Herder}, {DiPirro}, {Done}, {Dotani}, {Doty}, {Ebisawa}, {Enoto}, {Ezoe}, {Fabian}, {Ferrigno}, {Foster}, {Fujimoto}, {Fukazawa}, {Funk}, {Furuzawa}, {Galeazzi}, {Gallo}, {Gandhi}, {Gilmore}, {Guainazzi}, {Haas}, {Haba}, {Hamaguchi}, {Harayama}, {Hatsukade}, {Hayashi}, {Hayashi}, {Hayashida}, {Hiraga}, {Hirose}, {Hornschemeier}, {Hoshino}, {Hughes}, {Hwang}, {Iizuka}, {Inoue}, {Ishibashi}, {Ishida}, {Ishikawa}, {Ishimura}, {Ishisaki}, {Itoh}, {Iwata}, {Iyomoto}, {Jewell}, {Kaastra}, {Kallman}, {Kamae}, {Kataoka}, {Katsuda}, {Katsuta}, {Kawaharada}, {Kawai}, {Kawano}, {Kawasaki}, {Khangaluyan}, {Kilbourne}, {Kimball}, {Kimura}, {Kitamoto}, {Kitayama},
  {Kohmura}, {Kokubun}, {Konami}, {Kosaka}, {Koujelev}, {Koyama}, {Krimm}, {Kubota}, {Kunieda}, {LaMassa}, {Laurent}, {Lebrun}, {Leutenegger}, {Limousin}, {Loewenstein}, {Long}, {Lumb}, {Madejski}, {Maeda}, {Makishima}, {Markevitch}, {Masters}, {Matsumoto}, {Matsushita}, {McCammon}, {McGuinness}, {McNamara}, {Miko}, {Miller}, {Miller}, {Mineshige}, {Minesugi}, {Mitsuishi}, {Miyazawa}, {Mizuno}, {Mori}, {Mori}, {Moroso}, {Muench}, {Mukai}, {Murakami}, {Murakami}, {Mushotzky}, {Nagano}, {Nagino}, {Nakagawa}, {Nakajima}, {Nakamori}, {Nakashima}, {Nakazawa}, {Namba}, {Natsukari}, {Nishioka}, {Nobukawa}, {Noda}, {Nomachi}, {O'Dell}, {Odaka}, {Ogawa}, {Ogawa}, {Ogi}, {Ohashi}, {Ohno}, {Ohta}, {Okajima}, {Okazaki}, {Ota}, {Ozaki}, {Paerels}, {Paltani}, {Parmar}, {Petre}, {Pinto}, {Pohl}, {Pontius}, {Porter}, {Pottschmidt}, {Ramsey}, {Reis}, {Reynolds}, {Ricci}, {Russell}, {Safi-Harb}, {Saito}, {Sakai}, {Sameshima}, {Sato}, {Sato}, {Sato}, {Sawada}, {Serlemitsos}, {Seta}, {Shibano}, {Shida}, {Shimada}, {Shirron},
  {Simionescu}, {Simmons}, {Smith}, {Sneiderman}, {Soong}, {Stawarz}, {Sugawara}, {Sugita}, {Szymkowiak}, {Tajima}, {Takahashi}, {Takahashi}, {Takeda}, {Takei}, {Tamagawa}, {Tamura}, {Tamura}, {Tanaka}, {Tanaka}, {Tanaka}, {Tashiro}, {Tawara}, {Terada}, {Terashima}, {Tombesi}, {Tomida}, {Tsuboi}, {Tsujimoto}, {Tsunemi}, {Tsuru}, {Uchida}, {Uchiyama}, {Uchiyama}, {Ueda}, {Ueda}, {Ueno}, {Uno}, {Urry}, {Ursino}, {de Vries}, {Wada}, {Watanabe}, {Watanabe}, {Werner}, {White}, {Wilkins}, {Yamada}, {Yamada}, {Yamaguchi}, {Yamaoka}, {Yamasaki}, {Yamauchi}, {Yamauchi}, {Yaqoob}, {Yatsu}, {Yonetoku}, {Yoshida}, {Yuasa}, {Zhuravleva}, {Zoghbi}, \& {ZuHone}}]{hitomi_spie2014}
{Takahashi}, T., {Mitsuda}, K., {Kelley}, R., {et~al.} 2014, in Society of Photo-Optical Instrumentation Engineers (SPIE) Conference Series, Vol. 9144, Space Telescopes and Instrumentation 2014: Ultraviolet to Gamma Ray, ed. T.~{Takahashi}, J.-W.~A. {den Herder}, \& M.~{Bautz}, 914425

\bibitem[{{Turner} {et~al.}(2001){Turner}, {Abbey}, {Arnaud}, {Balasini}, {Barbera}, {Belsole}, {Bennie}, {Bernard}, {Bignami}, {Boer}, {Briel}, {Butler}, {Cara}, {Chabaud}, {Cole}, {Collura}, {Conte}, {Cros}, {Denby}, {Dhez}, {Di Coco}, {Dowson}, {Ferrando}, {Ghizzardi}, {Gianotti}, {Goodall}, {Gretton}, {Griffiths}, {Hainaut}, {Hochedez}, {Holland}, {Jourdain}, {Kendziorra}, {Lagostina}, {Laine}, {La Palombara}, {Lortholary}, {Lumb}, {Marty}, {Molendi}, {Pigot}, {Poindron}, {Pounds}, {Reeves}, {Reppin}, {Rothenflug}, {Salvetat}, {Sauvageot}, {Schmitt}, {Sembay}, {Short}, {Spragg}, {Stephen}, {Str{\"u}der}, {Tiengo}, {Trifoglio}, {Tr{\"u}mper}, {Vercellone}, {Vigroux}, {Villa}, {Ward}, {Whitehead}, \& {Zonca}}]{Turner2001}
{Turner}, M.~J.~L., {Abbey}, A., {Arnaud}, M., {et~al.} 2001, \aap, 365, L27

\bibitem[{Wang {et~al.}(2004)Wang, Bovik, Sheikh, \& Simoncelli}]{Wang_2004}
Wang, Z., Bovik, A., Sheikh, H., \& Simoncelli, E. 2004, IEEE Transactions on Image Processing, 13, 600

\bibitem[{{XRISM Science Team}(2020)}]{XRISM_white_paper2020}
{XRISM Science Team}. 2020, arXiv e-prints, arXiv:2003.04962

\bibitem[{Xu {et~al.}(2022)Xu, Huang, Deng, \& Yokoya}]{tensor}
Xu, T., Huang, T.-Z., Deng, L.-J., \& Yokoya, N. 2022, IEEE Transactions on Geoscience and Remote Sensing, 60, 1

\bibitem[{Yokoya {et~al.}(2017)Yokoya, Grohnfeldt, \& Chanussot}]{fusionreview}
Yokoya, N., Grohnfeldt, C., \& Chanussot, J. 2017, IEEE Geoscience and Remote Sensing Magazine, 5, 29

\bibitem[{Yokoya {et~al.}(2012)Yokoya, Yairi, \& Iwasaki}]{CNMF}
Yokoya, N., Yairi, T., \& Iwasaki, A. 2012, IEEE Transactions on Geoscience and Remote Sensing, 50, 528

\bibitem[{Zhang(2008)}]{zhang2008}
Zhang, Y. 2008, The International Archives of the Photogrammetry, Remote Sensing and Spatial Information Sciences, 37

\end{thebibliography}

\begin{appendix} 

\section{Metrics definition}
\label{sec:metrics}
In this section, we define the scalar metrics calculated in Table \ref{tab:error_metrics}. First, the Normalised Mean Squared Error (NMSE) is defined as such (where $Z_i$ is the value of $Z$ at the voxel $i$): 
\begin{equation}
    NMSE(\hat{Z},Z)=10\log_{10}\frac{\sum_{i=1}^{LMN}(\hat{Z}_i^2)}{\sum_{i=1}^{LMN}\Big((\hat{Z}_i-Z_i)^2\Big)},
    \label{eq:NMSE}
\end{equation}
By this definition, a higher NMSE implies a result of better quality. 
Then, the average Spectral Angle Mapper (aSAM) is defined as: 
\begin{equation}
    aSAM(\hat{Z},Z)=\frac{1}{L}\sum_{i}^{MN} \arccos \Bigg( \frac{\langle Z_i, \hat{Z}_i \rangle}{||Z_i||_2||\hat{Z}_i||_2} \Bigg),
    \label{eq:aSAM}
\end{equation}
where $Z_i$ is the spectra for the pixel $i$.
A lower aSAM implies a result of better quality, and an ideal value is 0.  

The average complementary Structural Similarity (SSIM) index, is:
\begin{equation}
    acSSIM(\hat{Z},Z)=1-\frac{1}{L} \sum_{k=1}^L \text{SSIM}(\hat{Z}_k,Z_k),
    \label{eq:acSSIM}
\end{equation}
where SSIM is the Structural Similarity index, as defined by \cite{Wang_2004}. The aim of the acSSIM is to measure the capacity to reconstruct local image structure, luminance, and contrast. The lower the acSSIM, the better the result. An ideal value is 0.

Finally, the Relative Dimensionless Global Error (ERGAS), is defined as: 
\begin{eqnarray}
    \text{ERGAS}(\hat{Z},Z,d)=100d\sqrt{\frac{1}{L}\sum_{k=1}^{L} \Bigg(\frac{||\hat{Z}_k-Z_k||_F}{\mu_k \sqrt{M N}}\Bigg)^2},
\end{eqnarray}

where $d$ is the ratio between the spatial resolution of $X$ and the spatial resolution of $Y$, and $\mu_k$ is the mean of $\hat{Z}_k$. The lower the ERGAS, the better, and an ideal value is 0. 

\section{Additional Figures}
\label{sec:add_figures}
\begin{figure*}[p]
    \centering
    \begin{subfigure}{\textwidth}
      \centering
    \includegraphics[width=\textwidth]{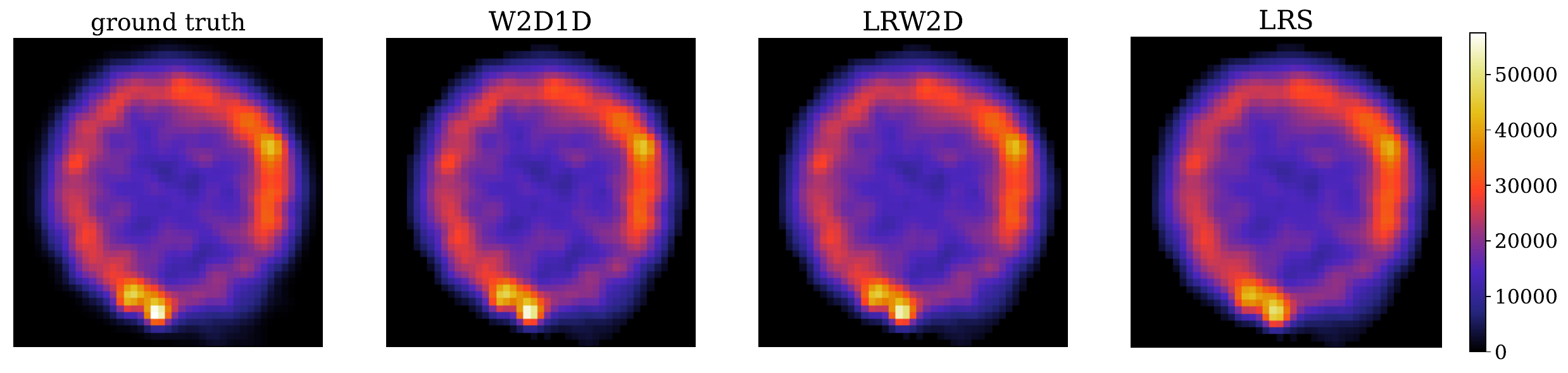}
    \caption{}
\end{subfigure}
\\
    \begin{subfigure}{\textwidth}
      \centering
    \includegraphics[width=\textwidth]{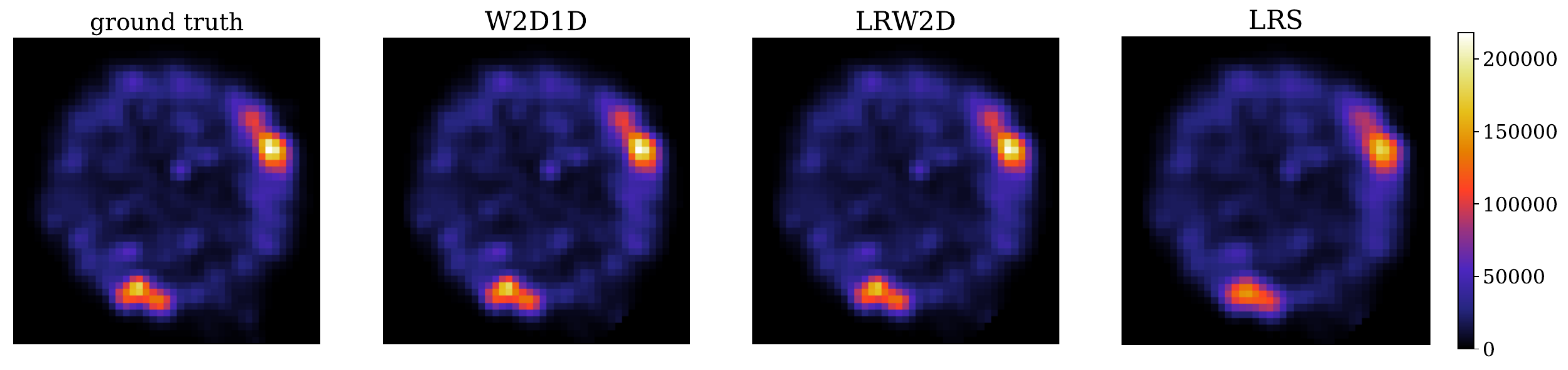}
    \caption{}
\end{subfigure}
\\
\begin{subfigure}{\textwidth}
      \centering
    \includegraphics[width=\textwidth]{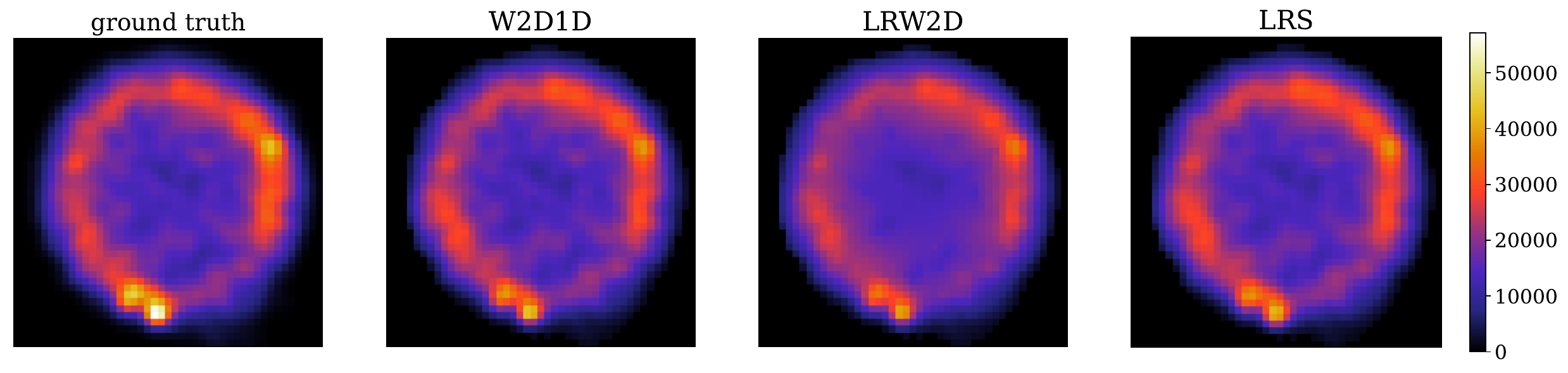}
    \caption{}
\end{subfigure}
\\
\begin{subfigure}{\textwidth}
      \centering
    \includegraphics[width=\textwidth]{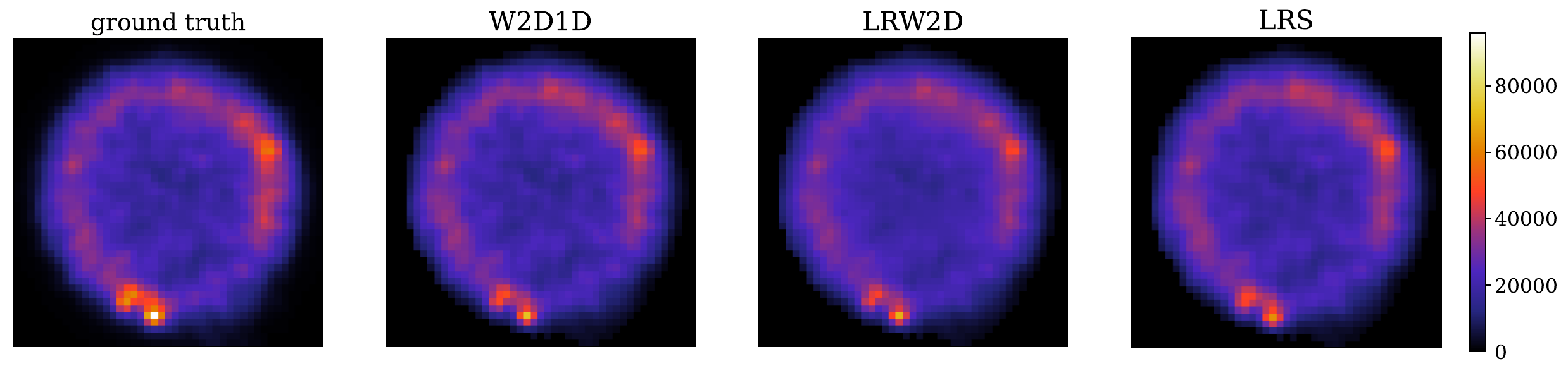}
    \caption{}
\end{subfigure}
\caption{Amplitude maps (i.e. sum of counts per pixel over every spectral channel) obtained by the three regularizers and compared to the ground truth.\\
(a) Gaussian model between 0.5-1.4 keV, (b) Gaussian model between 6.2-6.9 keV, (c) Gaussian model with rebinning between 0.5-1.4 keV, and (d), Realistic model between 0.5-1.4 keV. 
}
\label{fig:amp_maps}
\end{figure*}

\begin{figure*}
    \centering
    \begin{subfigure}{0.9\textwidth}
      \centering
    \includegraphics[width=\textwidth]{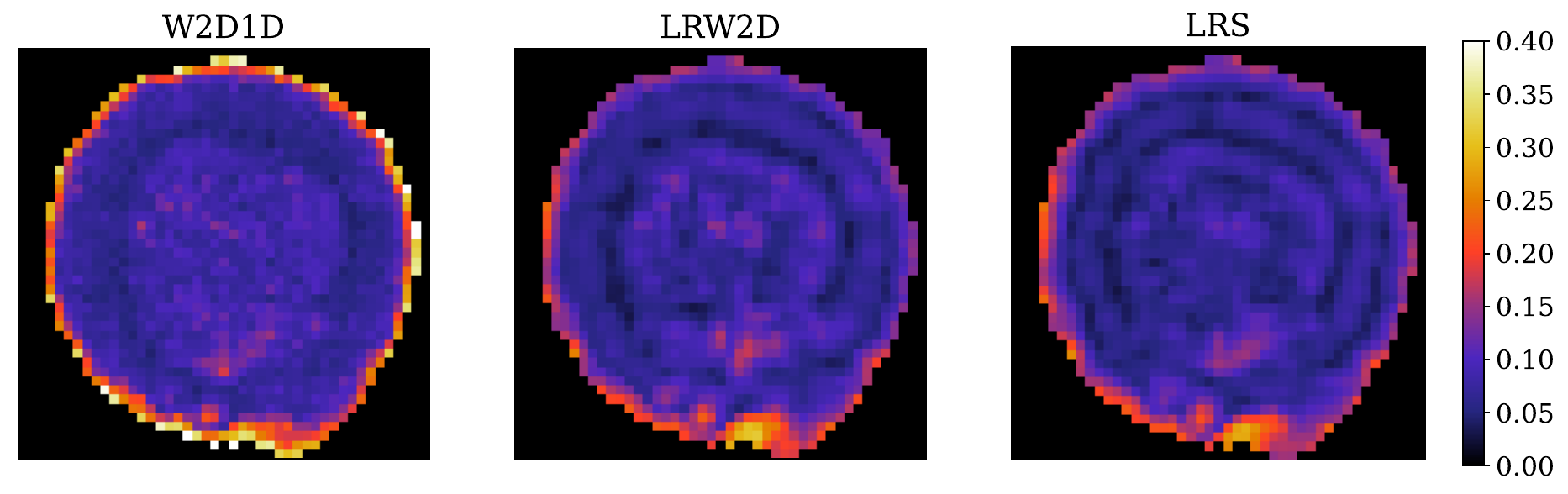}
    \caption{}
\end{subfigure}
\\
    \begin{subfigure}{0.9\textwidth}
      \centering
    \includegraphics[width=\textwidth]{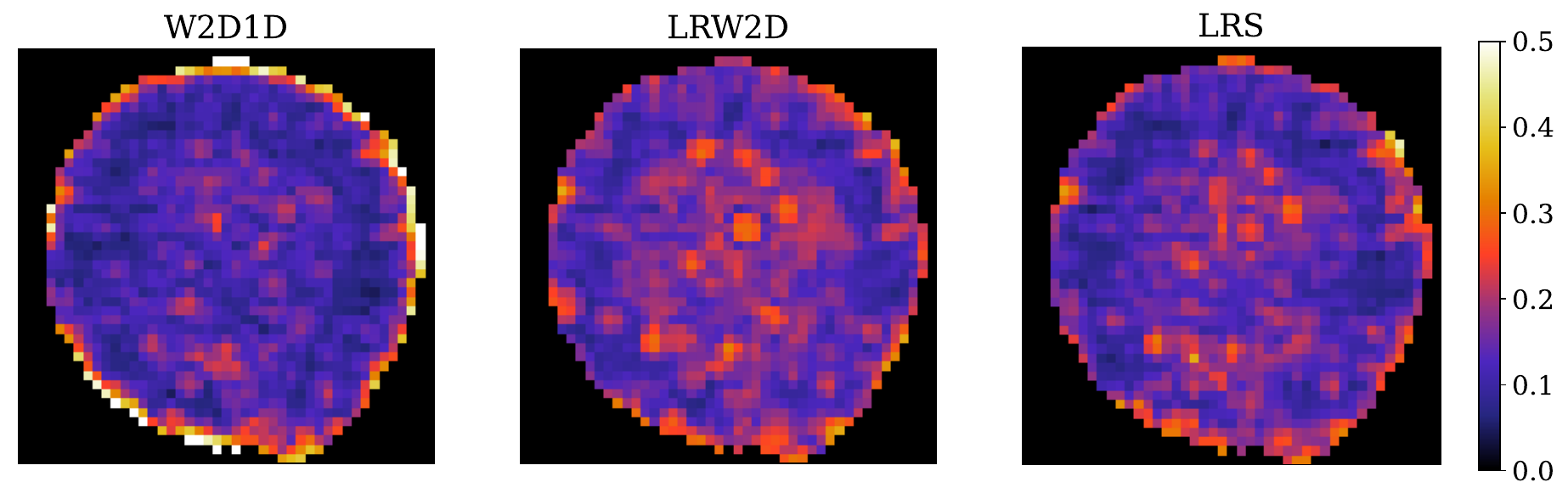}
    \caption{}
\end{subfigure}
\\
\begin{subfigure}{0.9\textwidth}
      \centering
    \includegraphics[width=\textwidth]{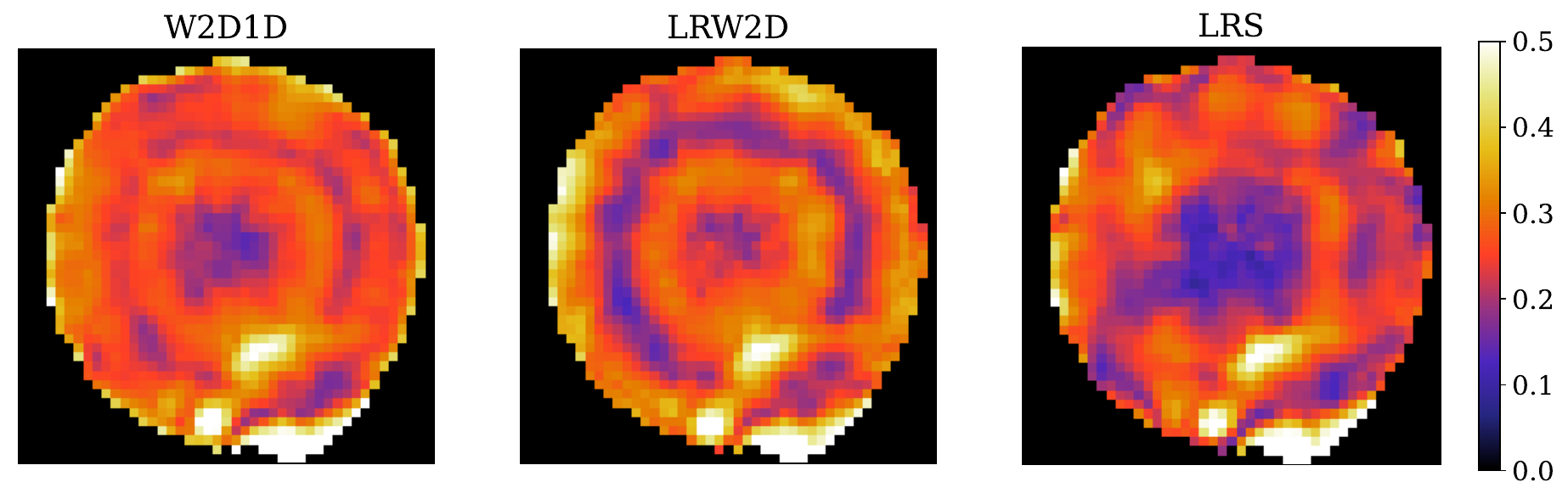}
    \caption{}
\end{subfigure}
\\
\begin{subfigure}{0.9\textwidth}
      \centering
    \includegraphics[width=\textwidth]{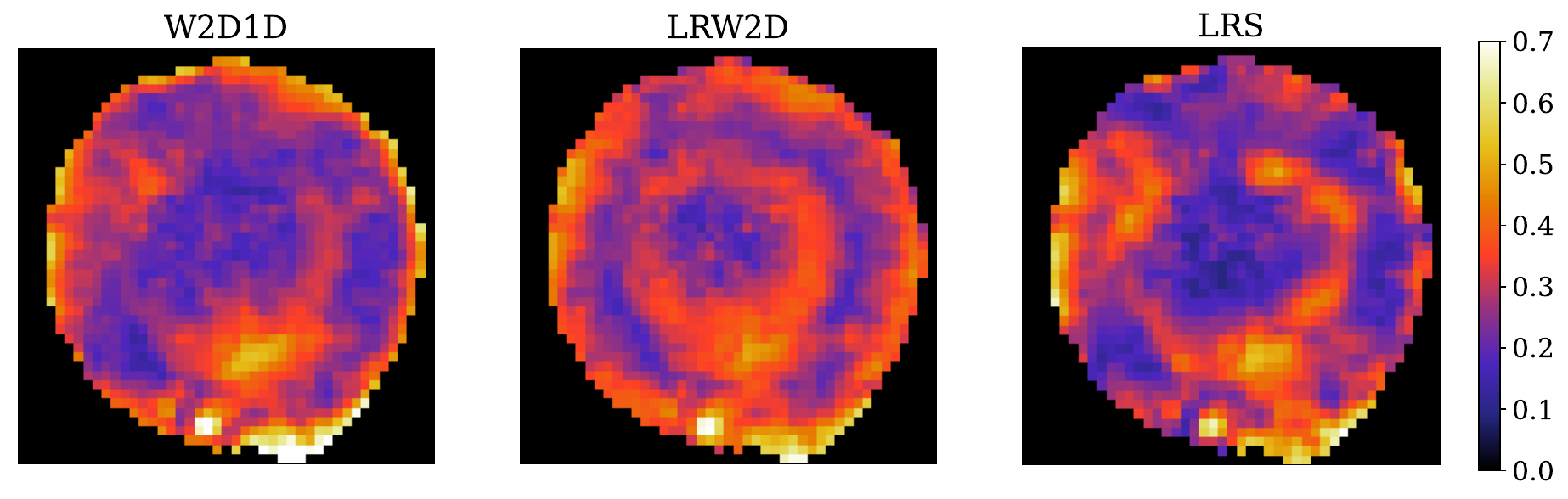}
    \caption{}
\end{subfigure}
\caption{Maps of the SAM (Equation \ref{eq:sam}) for the four toy models, and the three regularizations. A higher SAM means a worse spectral reconstruction. \\ 
(a) Gaussian model between 0.5-1.4 keV, (b) Gaussian model between 6.2-6.9 keV, (c) Gaussian model with rebinning between 0.5-1.4 keV, and (d), Realistic model between 0.5-1.4 keV. }
\label{fig:spectral_angle_err}
\end{figure*}

\begin{figure*}
    \centering
    \begin{subfigure}{0.9\textwidth}
      \centering
    \includegraphics[width=\textwidth]{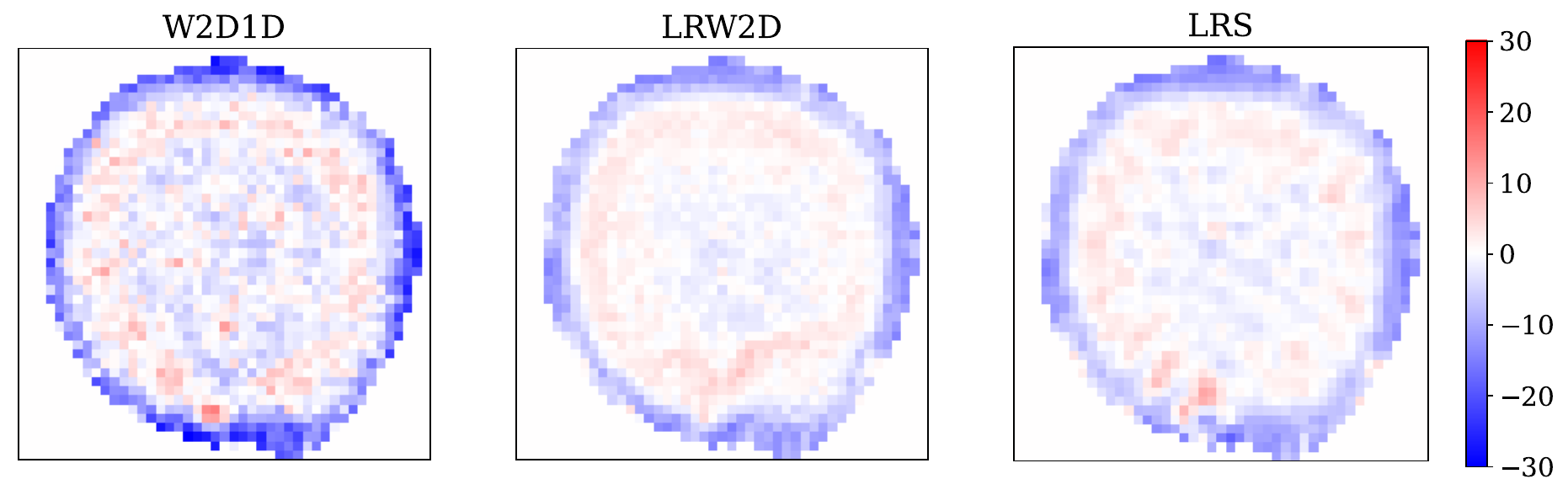}
    \caption{}
\end{subfigure}
\\
    \begin{subfigure}{0.9\textwidth}
      \centering
    \includegraphics[width=\textwidth]{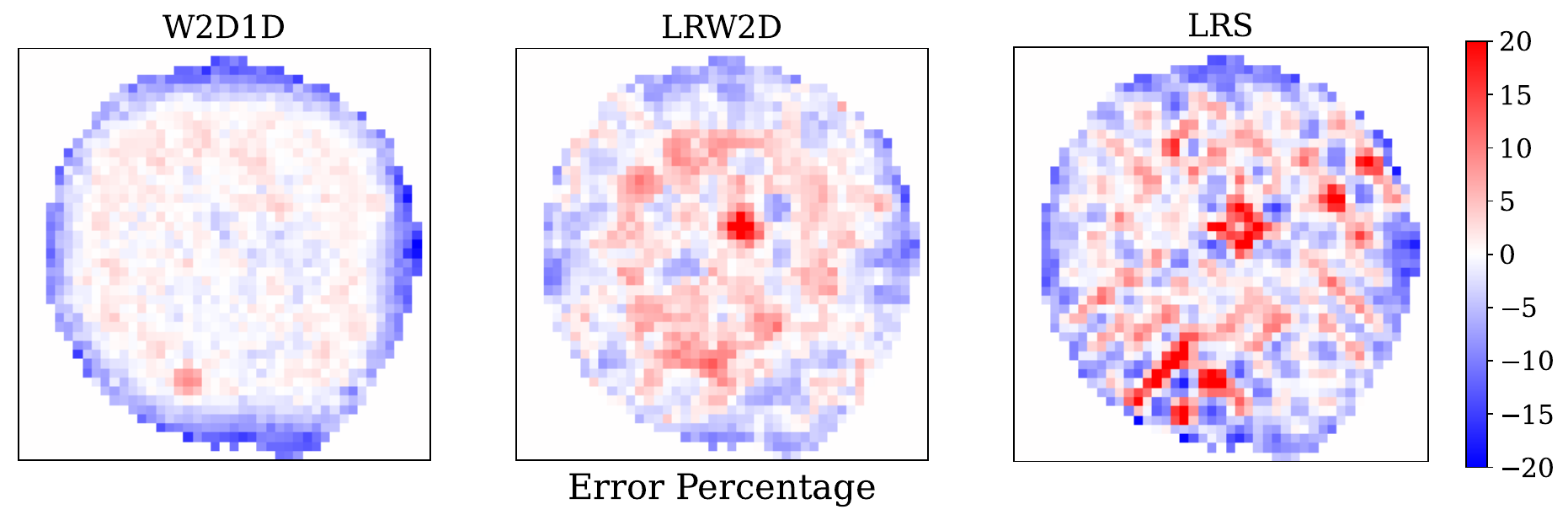}
    \caption{}
\end{subfigure}
\\
\begin{subfigure}{0.9\textwidth}
      \centering
    \includegraphics[width=\textwidth]{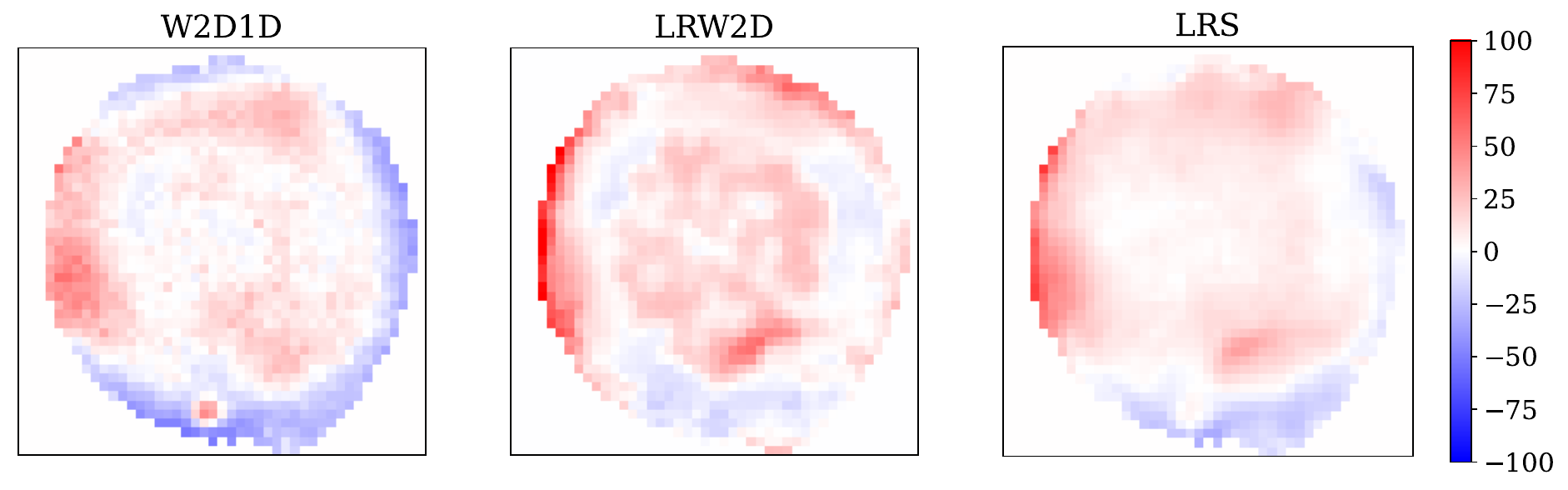}
    \caption{}
\end{subfigure}
\\
\begin{subfigure}{0.9\textwidth}
      \centering
    \includegraphics[width=\textwidth]{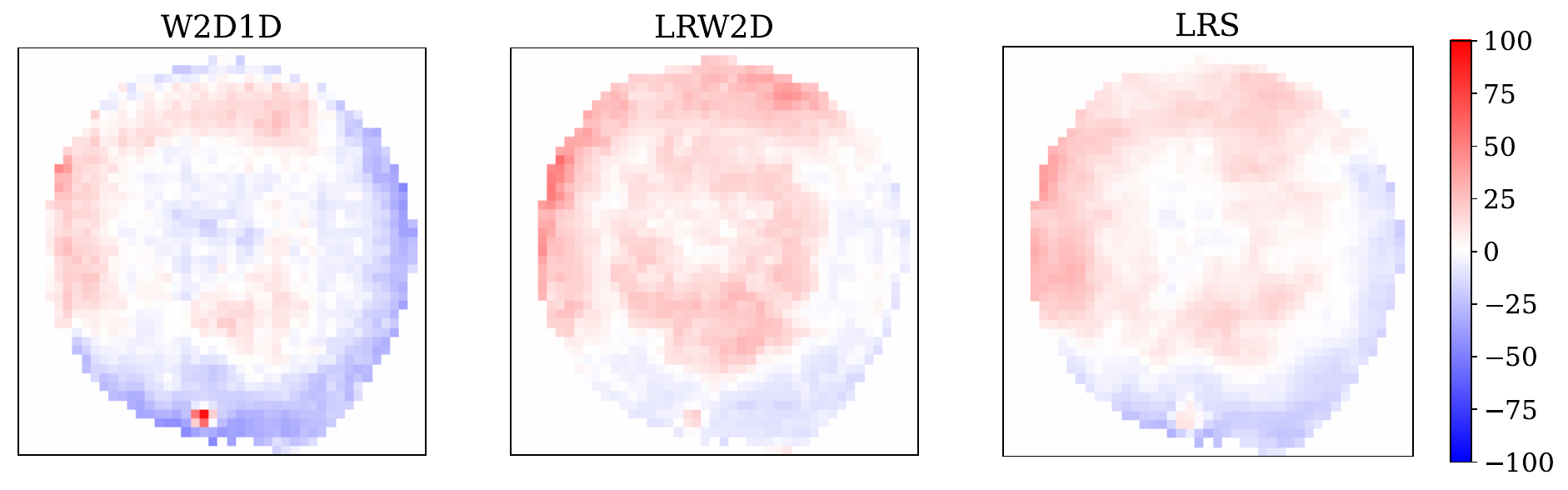}
    \caption{}
\end{subfigure}
\caption{Maps of the relative error: $Err(Z)=100(Z-\hat{Z})/\hat{Z}$, for the four toy models, and the three regularizations. A higher absolute error means a worse spectral reconstruction. \\
(a) Gaussian model between 0.5-1.4 keV, (b) Gaussian model between 6.2-6.9 keV, (c) Gaussian model with rebinning between 0.5-1.4 keV, and (d), Realistic model between 0.5-1.4 keV. }
\label{fig:err_percent}
\end{figure*}

This section presents additional figures for the results of section \ref{sec:results}. Fig. \ref{fig:amp_maps} shows the obtained amplitude for for the three potential regularization, for each of the four toy models. For these same options, Fig. \ref{fig:spectral_angle_err} shows the spectral angle maps.
Fig. \ref{fig:err_percent} displays the relative error maps averaged per pixel.

\end{appendix}


%

\end{document}